\documentclass[aps,prc,twocolumn,groupedaddress,showpacs,floatfix]{revtex4}
\usepackage{graphicx}
\usepackage{dcolumn}
\usepackage{longtable}
\usepackage{amsmath}
\begin{document}

\title{Nuclear Transparency in $90^\circ_{c.m.}$
 Quasielastic A(p,2p) Reactions}

\date{May 24, 2004}

\vspace{0.5cm}
\author{J.~Aclander$^{7,\dag}$,
J.~Alster$^{7,\dag}$,
G.~Asryan$^{1,b,\dag}$,
Y.~Averiche$^{5,\dag}$,
D.~S.~Barton$^{1,\ddag}$,
V.~Baturin$^{2,a,\dag}$,
N.~Buktoyarova$^{1,a,\dag}$,
G.~Bunce$^{1,\ddag}$,
A.~S.~Carroll$^{1,c,\ddag}$,
N.~Christensen$^{3,d,\dag}$,
H.~Courant$^{3,\ddag}$,
S.~Durrant$^{2,\dag}$,
G.~Fang$^{3,\S}$,
K.~Gabriel$^{2,\dag}$,
S.~Gushue$^{1,\ddag}$,
K.~J.~Heller$^{3,\S}$,
S.~Heppelmann$^{2,\ddag}$,
I.~Kosonovsky$^{7,\dag}$,
A.~Leksanov${^2,\dag}$,
Y.~I.~Makdisi$^{1,\ddag}$,
A.~Malki$^{7,\dag}$,
I.~Mardor$^{7,\dag}$,
Y.~Mardor$^{7,\dag}$,
M.~L.~Marshak$^{3,\ddag}$,
D.~Martel$^{4,\dag}$,
E.~Minina$^{2,\dag}$,
E.~Minor$^{2,\dag}$,
I.~Navon$^{7,\dag}$,
H.~Nicholson$^{8,\dag}$,
A.~Ogawa$^{2,\dag}$,
Y.~Panebratsev$^{5,\dag}$,
E.~Piasetzky$^{7,\dag}$,
T.~Roser$^{1,\dag}$,
J.~J.~Russell$^{4,\ddag}$,
A.~Schetkovsky$^{2,a,\dag}$,
S.~Shimanskiy$^{5,\dag}$,
M.~A.~Shupe$^{3,e,\S}$,
S.~Sutton$^{8,\dag}$,
M.~Tanaka$^{1,f,\dag}$,
A.~Tang$^{6,\dag}$,
I.~Tsetkov$^{5,\dag}$,
J.~Watson$^{6,\dag}$
C.~White$^{3,\dag}$,
J-Y.~Wu$^{2,\dag}$,
D.~Zhalov$^{2,\dag}$
}

\affiliation{
\mbox{$^1$~Brookhaven~National~Laboratory,~Upton,~NY~11973-5000,~USA} \\ 
\mbox{$^2$~Pennsylvania~State~University,~University~Park,~Pennsylvania~16802,
~USA} \\
\mbox{$^3$~University~of~Minnesota,~Minneapolis,~Minnesota~55455,~USA} \\
\mbox{$^4$~University~of~Massachusetts
~Dartmouth,~North~Dartmouth,~Massachusetts~02747,~USA} \\
\mbox{$^5$~J.I.N.R.,~Dubna,~Moscow~141980,~Russia} \\
\mbox{$^6$~Dept.~of~Physics,~Kent~State~University,~Kent,~Ohio~44242,~USA} \\
\mbox{$^7$~School~of~Physics~and~Astronomy,~Tel~Aviv~University,~Tel~Aviv
~69978,~Israel} \\
\mbox{$^{8}$~Dept.~of~Physics,~Mount~Holyoke~College,~South~Hadley,
~Massachusetts~01075,~USA} \\
\mbox{$^a$~Home~Institution~is~Petersburg~Nuclear~Physics~Institute,~Gatchina,
~St~Petersburg~188350,~Russia} \\ 
\mbox{$^b$~Home~Institution~is~Yerevan~Physics~Institute,~Yerevan~375036,
~Armenia} \\  
\mbox{$^c$~Now~at~71~Kendal~Drive,~Oberlin,~Ohio~44074,~USA} \\ 
\mbox{$^d$~Now~at~Carleton~College,~Northfield,~Minnesota~55057,~USA} \\ 
\mbox{$^e$~Now~at~University~of~Arizona,~Tucson~Arizona~85721,~USA} \\ 
$^f$~deceased \\ 
$^\S$~Member of AGS Experiment 834. \\
$^\dag$~Member of AGS Experiment 850. \\
$^\ddag$~Member of AGS Experiments 834 and 850. \\
}

\begin{abstract}
We summarize the results of two experimental programs at the Alternating 
Gradient Synchrotron of BNL to measure the nuclear transparency of nuclei
 measured in the A(p,2p) quasielastic scattering process near $90^\circ$
in the pp center of mass.  The incident momenta varied from 5.9 to 14.4 GeV/c,
corresponding to $4.8 < Q^2 < 12.7 (GeV/c)^2 $.
Taking into account the motion of the target proton in the nucleus, the 
effective incident momenta extended from 5.0 to 15.8 GeV/c.
 First, we describe the measurements with the newer
experiment, E850, which had more complete kinematic definition of quasielastic 
events.
E850 covered a larger range of incident momenta, and thus provided more
information regarding the nature of the energy dependence of the 
nuclear transparency.
 In E850 the angular dependence of the nuclear transparency near  $90^\circ$,
 and the nuclear transparency for deuterons was studied.
 Second, we review the techniques used in an earlier experiment,
E834, and show that the two experiments are consistent for the Carbon data.
E834 also determines
the nuclear transparencies for lithium, aluminum, 
copper, and lead nuclei as well
as for Carbon.  A determination of the ($\pi^+,\pi^+ p$) 
transparencies is also reported.
We find for both E850 and E834 that the A(p,2p) nuclear transparency, unlike
that for A(e,e'p) nuclear transparency,
 is incompatible with a constant value versus energy as
predicted by Glauber calculations.  The A(p,2p) nuclear transparency for
Carbon and Aluminum increases by a factor of two between 5.9 and 9.5 GeV/c 
incident proton momentum.  At its peak the A(p,2p) nuclear transparency
is $\sim80\%$ of the constant A(e,e'p) nuclear transparency.
Then the nuclear transparency falls back to a value
at least as small as that at 5.9 GeV/c, and is compatible with the Glauber
level again.  This oscillating behavior is generally interpreted as an 
interplay between two components of the pN scattering amplitude; one 
short ranged and perturbative, and the other long ranged and strongly 
absorbed in the nuclear medium.
A study of the A dependent nuclear transparency indicates that the effective 
cross section varies with incident momentum and is considerably smaller
than the free pN cross section.
We suggest a number of experiments for further studies of nuclear
transparency effects.
 
  
\end{abstract}

\maketitle
\section{Introduction}

\begin{figure}[hbp]
\includegraphics[width=65mm]{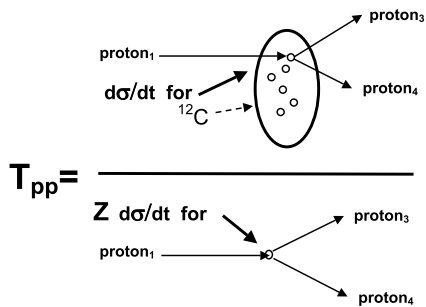}
\caption{Illustration of the quantities used in the determination 
of nuclear transparency for a representative nucleus, $^{12}C$ in the 
numerator, for free pp scattering in the denominator.   
  The incident proton is $proton_1$, and the struck proton
is $proton_2$.   The two outgoing protons are designated as $proton_3$ and 
$proton_4$. }
\label{fig:Tr_illust}
\end{figure}

If the nucleons in a nucleus were at rest and very lightly bound,
 then nuclear transparency for A(p,2p) reactions
 as illustrated in Figure~\ref{fig:Tr_illust}
would simply be the ratio of the  differential cross section for 
quasielastic scattering from the protons in the nucleus (e.g. Carbon),
to the 
differential cross section for free pp scattering corrected for 
 the number of protons in the nucleus, Z.  The nuclear transparency is then
a measure of the survival probability for the protons to enter
 and exit the nucleus without interacting with the spectator nucleons in the
target nucleus.
The actual situation is significantly complicated by the 
momentum and binding energy distributions described by the
spectral function of the protons in the nucleus. Note that in this paper
we will be implicitly integrating the spectral functions over the 
 binding energy distributions and considering only the nuclear momentum 
distributions. 
Even with the assumption that the scattering can be factorized,  
 a detailed knowledge of the behavior of the
elementary pp differential cross section, and of the spectral
function of the nucleus is required since the pp differential cross section at 
large angles depends so strongly on energy.
 In the experiments described below, 
either a sample of protons with a narrow range
 longitudinal momentum is selected, or
the observed quasielastic distributions are corrected 
with the known differential pp cross section.
Similar transparencies can be defined for other quasielastic
scattering processes by other incident hadrons or leptons.
  Transparencies for  A(e,e'p)  have  been extensively measured as
discussed below .

In this paper we report on the combined results of two AGS experiments to 
measure the 
nuclear transparency of nuclei 
in the A(p,2p) quasielastic scattering process near $90^\circ$ 
in the pp center of mass (c.m.).
In the first part of the paper we describe the measurements with the newer
experiment, E850, which features a  more complete kinematic definition of the 
quasielastic events. E850
extends the range of incident energies, and provides more information
regarding  the nature of the
unexpected fall in the nuclear transparency above 9.5 GeV/c 
\cite{IM},\cite{E850}.
In the second part, we review the techniques used in the earlier  E834 
experiment \cite{E834}.
For the E834 experiment, the directions of each final state particle were 
determined
but the momentum of only one track was measured.  The E850  experiment allows 
full and symmetrical
tracking with momentum reconstruction of both final state particles.  The E850 
measurement addresses 
the concerns about the background subtraction in the determination of the 
quasielastic signal in the
E834 experiment.  We show for the overlapping Carbon measurements 
that these two experiments
are consistent in the energy dependence of the nuclear transparency.
E834 measures the nuclear transparency for five different nuclei, and also
 yields initial results regarding transparencies for A$(\pi^{+},\pi^{+}p)$
 interactions \cite{JYW}.  While most of the 
principal results have been reported previously, we present a 
more detailed and consistent view of the two programs.

        Other measurements with these two detectors allow us to investigate
 a number of the factors involved with the measurement and interpretation of
 nuclear transparency.  The publication of Y.~Mardor, {\it et al.} reported
on a study of the factorization assumption, and the equality of 
the longitudinal and transverse portions of the nuclear momentum distribution
  for the Carbon nucleus
\cite{Mardor:1998fz}.  Short range correlations, which give rise to the high 
momentum tails of the momentum distributions were reported
 in papers by Aclander,
 {\it et al.}, Tang, {\it et al.} , and Malki, {\it et al.}
   \cite{Aclander:1999fd,Tang,Malki:2000gh}.  A(p,2p) measurements from both
 the E834 detector by Heppelmann, {\it et al.}  \cite{SpecFn}, and the
 E850 detector by Y.~Mardor,  {\it et al.} \cite{Mardor:1998fz}
    showed that the nuclear momentum distributions
 determined in A(p,2p) reactions were in agreement with
 those found by A(e,e'p)  experiments  .

Color transparency refers to a QCD phenomenon, independently
 predicted in 1982 by Brodsky \cite{SB} and 
Mueller \cite{MB}, involving a reduction of soft interactions, in both the 
initial (ISI) 
and final (FSI) states, for a hard quasielastic scattering. 
These theorists deduced from QCD that when a proton traversing 
the nucleus experiences a hard collision, a special quantum 
state is selected. That special state involves the 
part of the proton wave function that is most `shock resistant'
and  tends to survive the hard collision 
without breaking up or radiating  gluons. This state 
is also expected to survive long enough while traveling through the
nucleus to  
have a reduced interaction with the 
spectators in the target nucleus. The state is predicted 
to involve a rare component of the proton wave function 
that is dominated by 3 valence quarks at small transverse 
spatial separation. The color transparency prediction of
QCD is that the fraction of nuclear protons contributing 
to A(p,2p) quasielastic scattering should increase from a 
nominal level consistent with Glauber absorption \cite{RG,FSZ},  
at low $Q^2$, and then to approach unity at very high  $Q^2$. 
However, the fall of nuclear transparency above 9.5 GeV/c 
indicates that additional amplitudes are required.

Due to the very steep dependence of the $90^\circ$ pp cross section on 
the center of mass energy squared, $s^{-10}$, and the uncertainties
in the nuclear momentum distributions, it is useful to
measure a ratio close to the kinematic point where the target proton
is at rest, particularly in the longitudinal direction.
  This ratio will demonstrate the energy dependence
of the nuclear transparency with a minimum of assumptions.
We refer to this as the nuclear transparency ratio, $T_{CH}$,
for quasielastic scattering on Carbon, compared to that
for hydrogen.  Then using our best knowledge of the energy dependence of the 
pp cross section, and the nuclear momentum distributions we
can determine the nuclear transparency, $T_{pp}$, integrated in the
entire longitudinal direction.

There have been a number of investigations of nuclear transparency in addition
to those involving quasielastic reactions.  Transparencies in exclusive
incoherent $\rho^0$ production have been measured at Fermilab, CERN, and DESY.
At Fermilab \cite{Fermi_mu_rho}, increases in exclusive nuclear transparency 
have
been measured as the photon becomes more virtual, as expected in the color
transparency picture.  CERN data, also involving muon production of $\rho^0$'s,
but at higher $Q^2$, indicate that the effect is smaller \cite{CERN_mu_rho}. 
Coherence
length effects have been investigated in the HERMES experiment at DESY
\cite{HERMES_e_rho}. It is important to distinguish between coherence length, 
which
is the distance at which the virtual photon fluctuates into a $q\overline{q}$ 
pair,
and the formation length in which the initially point like $q\overline{q}$ 
pair grows to the
normal $\rho^0$ size.  It is the formation length that enters into the
determination of color transparency.  The coherence length is not 
a factor in quasielastic scattering.

Another important investigation of nuclear transparency has been the coherent
diffractive dissociation of 500 GeV/c $\pi^-$ into di-jets on nuclear
targets at $Q^2 > 7 (GeV/c)^2$  \cite{di-jet}.  Unlike the other experiments, 
di-jet production is not an exclusive reaction, 
and therefore its interpretation
may be different than other searches for color transparency.  The power
law behavior with A 
 determined from the A-dependence in this experiment is considerably
 larger, $\sim1.5$ than  the $\sim0.7$ usually found in inclusive $\pi$A
 reactions. The very high energy of this experiment 
makes color transparency effects
 likely, but does not allow a study of the threshold for nuclear transparency.

 This paper will confine itself to only quasielastic reactions, which have 
involved incident hadrons, p and $\pi^+$, and those 
with incident electrons.  The mechanisms of
 these quasielastic reactions are closely related, 
and therefore can be compared in a relatively straightforward manner.

\section{Experiment E850}

This section will cover measurements of nuclear transparency with
the E850 detector during two experimental runs. Details of the
detector characteristics, and  of the kinematical analysis are
discussed.

\subsection{The E850 Detector (EVA)}

\begin{figure}[htbp]
\includegraphics[width=85mm]{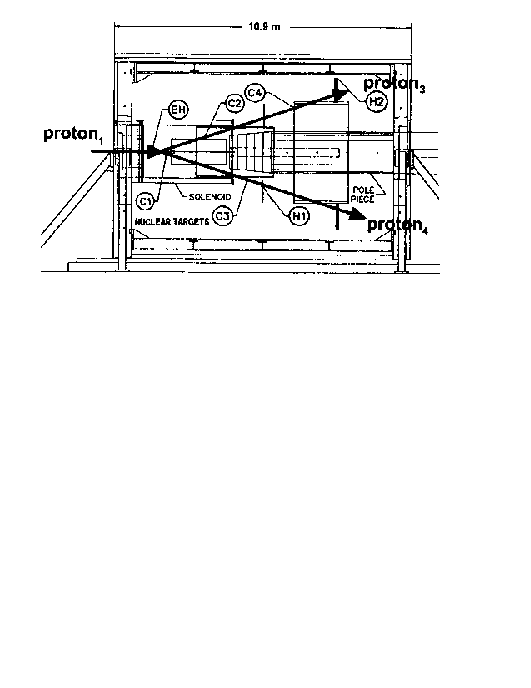}
\caption{Schematic Drawing of E850 solenoidal detector, which shows 
a vertical, midplane section. C1-C4 are the four-layer
arrays of straw tube cylinders. C1 and C2 were completely within the solenoid, 
while C3 was only partially inside the solenoid. 
C, $\text{CH}_2$, and $\text{CD}_2$ targets were located inside the C1  
straw tube cylinder.  The counter-weighted pole piece returns the magnetic
flux emerging from the right hand end of the solenoid. 
A re-entrant cavity in the
pole piece absorbs the incident beam downstream of the detecting elements.
   H1 and H2 were the scintillation
trigger counters of about 256 elements each.
  The bold arrows show typical trajectories of the incident
particle, $proton_1$ and two scattered protons, 
$proton_3$ and $proton_4$. 
For scale, the length of the magnet frame was 10.9m, and the straw tube
cylinders (C2, C3, and C4) were 2m long.
}
\label{fig:solenoid}
\end{figure}

A 24 GeV proton beam from the AGS produced secondary hadrons at $0^\circ$
 through its
interaction with a $\sim3$ cm long platinum target.  After magnets in the C1
transport line dispersed the secondary hadrons with respect to charge
 and momentum, collimators selected the particles to be transported to the
experiment. 
 The momentum spread of the beam was typically $\pm0.5\%$.
Every 3 seconds, up to $5{\times}10^7$ particles in a 1 second long spill 
were delivered to a 1$\times$2 $cm^2$ spot.   
Two, pressurized $\text{CO}_2$, differential \v{C}erenkov counters,
 $\sim30$ m upstream
of the detector, allowed identification
of the pions, kaons and protons in the incident beam 
on a particle-by-particle basis \cite{Kycia,CW}. The incident particles 
were tracked by a series of scintillation hodoscopes along the beam line.
We measured the  incident beam flux with an ion chamber cross-calibrated 
with direct particle counting at lower intensities, and then corrected for the
proton fraction with the differential \v{C}erenkov counters.
The final hodoscope was the one denoted by BH just before
the nuclear targets in Figure~\ref{fig:solenoid}.

The E850 experiment embedded the Carbon (C), $\text{CH}_2$, and $\text{CD}_2$ 
targets inside a 2 $m$ 
diameter, 3 $m$ long
 superconducting solenoid with a 0.8 $T$ field as shown in 
Figure~\ref{fig:solenoid}. The solenoid was a modification 
the CLEO I solenoid originally used at Cornell \cite{CLEO_1}. 
The pole piece intercepts most of the magnetic flux emerging from 
the solenoid
to form a reasonably uniform, horizontal magnetic field through out the
solenoid volume.  The precise shape of the
field was determined by a combination of a large number
of Hall probe measurements in regions of high gradients and  3-D mesh 
calculations \cite{meng}.
  The annular space between the
pole piece and the solenoid allowed particles 
scattered near $90^\circ_{c.m.}$ to 
reach the detectors outside the solenoid.  This arrangement facilitates easier
triggering and better momentum resolution than having all the detector 
elements inside
the solenoid. The targets were 5.1 $\times$ 5.1 $cm^2$ in area, and
6.6 $cm$ in length, spaced $\sim20$ $cm$ apart.  The positions of the
C, $\text{CH}_2$and $\text{CD}_2$ targets were  interchanged regularly to
 minimize the differences in flux and acceptance.

   Surrounding the targets were four concentric cylinders 
(C1-C4) of mean radii: 10, 45, 90, and 180 cm.
Each cylinder was fabricated from four layers of thin wall straw tubes 
(114 $\mu$m Mylar with 8 $\mu$m
of aluminum cladding except for C4, which had double the wall
thickness).  Their diameters were  0.50, 1.04, 1.04, and 2.16 cm,
and their  total number was  5632 \cite{Kmit}. C1 and C2 were complete
cylinders, but C3 and C4 had small gaps ($\sim2\%$ and $\sim5\%$ in the 
azimuthal coverage) at the top and bottom to allow for support structures.
They were filled with 
a 50-50 mixture of Argon-Ethane at slightly above atmospheric pressure.
 All the tubes measure the drift distance
from the track to the central wire with an accuracy of ~150 $\mu$m, 
and for C2-C4, the longitudinal distance
was measured to $\sigma \simeq 2 cm$  by charge division 
\cite{Wu},\cite{Fernando}.    
The momentum resolution was dominated by multiple scattering, and was  
$\sigma(p)/p \simeq7\%$
as determined from pp elastic scattering.
\begin{figure}[htbp]
\includegraphics[width=45mm]{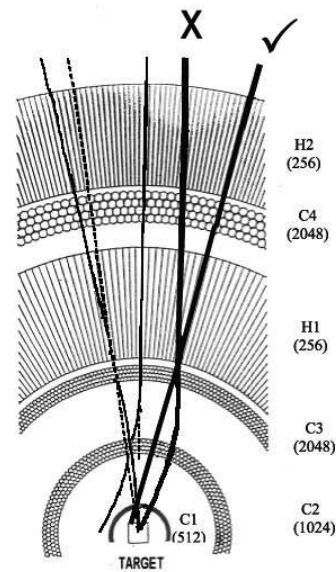}
\caption{A transverse (r-$\phi$) projection of a portion of the spectrometer 
(EVA) 
used for E850.  C1-C4 are cylinders of straw tubes, and H1 and H2 are hodoscope
arrays of scintillation counters.  The relative sizes of the elements
are only approximate. 
 Radially, the coil of the superconducting
solenoid just surrounds the  C3 cylinder, and the magnetic field
is nearly confined within this radius. The number of elements listed for each
array is nominal.  A few elements are missing in top and bottom in 
C3, C4 and H2 to allow for support structures.  C1, C2 and H1 cover the
entire azimuth.   The elements of H2 were 
one-third overlapped to provide twice the resolution in $\phi$.
The curved track labeled with the $\times$ was rejected by the trigger system, 
while the higher momentum, straighter track indicated by the $\surd$ was 
accepted.
}
\label{fig:E850_proj}
\end{figure}
The tracks scattered near $90^\circ$ c.m. passed through the 
annulus between the solenoid and the steel
pole piece until they reached the two fan-shaped scintillation counter arrays,
H1 and H2, and the two larger  straw tube cylinders, C3 
and C4.

As illustrated in Figure~\ref{fig:E850_proj} the trigger system selects only 
events with particles 
above a minimum transverse momentum, $P_T$ \cite{Wu}. 
This selection was done in two 
stages.   The first selection was 
done in $\sim75$ nsec by checking the correlation in $\phi$ between H1 and H2. 
Then a second more 
precise selection was performed by logic arrays, which 
measured the momentum by a 3-point correlation with the hits in the cylinders
labeled C2, C3 and C4.  
The time that this calculation took was  $\sim1\mu\text{sec}$, but
depended on the complexity of the event.
Some further checks were carried out by microprocessors attached to
 this trigger 
system to insure that there were two tracks and they were very 
crudely coplanar.
A microprocessor
then read out the CAMAC based TDC's and the FASTBUS based ADC's from the straw 
tubes and scintillation
counters.  Details of these trigger systems can be found in the 
reference of Wu, 
et al \cite{Wu}.  
\begin{figure}[htbp]
\includegraphics[width=85mm]{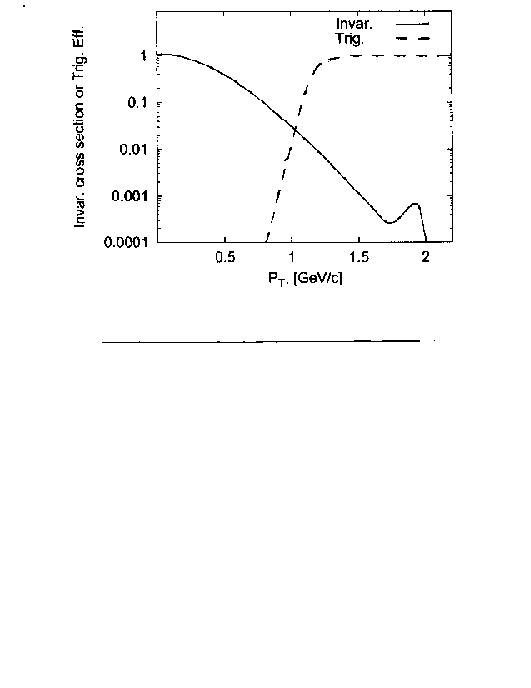}
\caption{The 8 GeV/c  invariant cross section at $90^o_{c.m.}$ 
for $pp \rightarrow p + X$, normalized to 1 at $P_T = 0$,
is shown as the solid line \cite{Ray_Th}.
The peak at the end is a representation of elastic scattering contribution.
 An approximate E850 
trigger acceptance is shown as the dashed line. 
}
\label{fig:trig_eff}
\end{figure}
Figure~\ref{fig:trig_eff} shows the inclusive pp cross section at $90^{\circ}$ 
for
8 GeV/c incident protons, and the approximate trigger efficiency, 
which rejects low $p_T$ inelastic backgrounds while retaining 
all of the 
events near the exclusive limit.  This arrangement provided an acceptable 
trigger rate of less than 100 Hz for 
incident beams of up to $10^8$ Hz ($\sim10^7$ interactions per spill). 

In a large detector such as EVA, the location, identification, and measurement
of tracks is a complex matter that will only 
be described  briefly in this article.  For more
detailed information, see the theses of References~ 
\cite{Simon_th},\cite{Yael_th},\cite{Israel_th},\cite{Alex_th},and 
\cite{Daniel_th}.
The initial location of all the straw tube elements was determined using a 
precision
3-D survey using the computer linked ManCat (surveying)
 system \cite{Mancat}.
This achieved a precision of better than 1 mm, which was further reduced 
to $\sim150\mu$m by fitting to magnet-off, straight tracks.

Legitimate hits in the straw tubes were selected by checking that the drift 
times were within physical limits, 
and that the ADC values were above the noise level.
Neighboring hits within  the 4 layers of each cylinder were collected 
into `bunches' of typically 3 or 4 hits.  Then these `bunches' were
combined with the `bunches' from the other cylinders to form complete tracks.
A multi-step process was used to select the best hits in the case of 
ambiguities, and
find the best fit to the tracks' curvature in the magnetic field. 
The sagitta of these tracks in the $r-\phi$ plane between
C1 and C3, and the deviation from the straight
line extended inward from hits in C3 and C4
were used to determine the particles'
momentum.   Charge division as measured by the
ADC values read from the two ends of the straw tubes 
 determined the z-coordinates of the straws 
selected in the $r-\phi$ fit.
These z coordinates in turn were fit to calculate the
polar angles of the tracks.
Final selections of events were the result of an 
adequate $\chi^2$ to the $r-\phi$ fit, and
a consistency of the vertex position as determined by the two tracks. 


\subsection{Kinematics for E850}
\label{subsec:E850_kinem}
In this experiment, the exclusive quasielastic process 
involves a single hard pp interaction with $Q^2 > 4(GeV/c)^2$.
The final state consists of two energetic protons and a residual excited 
nucleus. The presence of more than two tracks in the detector
identified the class of events with primary and secondary
inelastic interactions.  This allowed for background subtraction as
described below.

The fundamental sub-process of our quasielastic 
events is the pp interaction
\begin{equation}
proton_1 + proton_2 \rightarrow proton_3 + proton_4 \nonumber
\label{proton}
\end{equation}
at a pp scattering angle near  $90^\circ$ c.m.. We associate 
$proton_1$ with the beam particle,  and $proton_2$
with  the target 
proton in the nucleus. The quasielastic events are 
characterized by a small missing energy, $\epsilon_m$,
 and missing momentum, $\vec{P}_m$. 
We define missing energy, missing momentum and missing mass squared, $m_{M}^2$,
in terms of measured {\bf$P_i$}, and resulting energy, $E_i$  of each 
$proton_i$:
\begin{eqnarray}
 \epsilon_m = E_3 + E_4 - E_1 - m_p \nonumber \\       
 \stackrel{\rightarrow}{P}_m = \stackrel{\rightarrow}{P}_3 + 
\stackrel{\rightarrow}{P}_4 - \stackrel{\rightarrow}{P}_1   \nonumber \\
 m_{M}^2 = \epsilon^2_m -  \stackrel{\rightarrow}{P^2}_m. 
\label{eqn:P_miss}
\end{eqnarray}

In the spirit of the impulse approximation we identify the measured 
missing momentum with the momentum of the struck proton inside the nucleus.
For quasielastic scattering at c.m. scattering angles of $90^\circ$, the 
pair of final state protons is produced at approximately 
equal momenta, polar angles and opposite azimuth angles.
The simple symmetrical nature of the 
final state is altered by three classes of phenomena: 
the small variation in c.m. scattering angles around 
90$^\circ$; 
the effects of the motion of the struck proton determined by the nuclear 
momentum distribution 
and the interaction of initial or final state protons with the 
spectator nucleons in the nucleus.

By including a range of c.m. scattering angles around 90$^\circ$,
one induces a spread in the final state energies and polar angles, 
which are  proportional to
the $cos(\theta_{c.m.})$ \cite{IM}. 
The detector is configured to have acceptance
for c.m. scattering angles in the range  
$\sim 86^\circ < \theta_{c.m.} \leq 90^\circ$
at each beam energy as listed in Table~\ref{tab:E850_T} of the Appendix.

\begin{figure}[htbp]
\includegraphics[width=85mm]{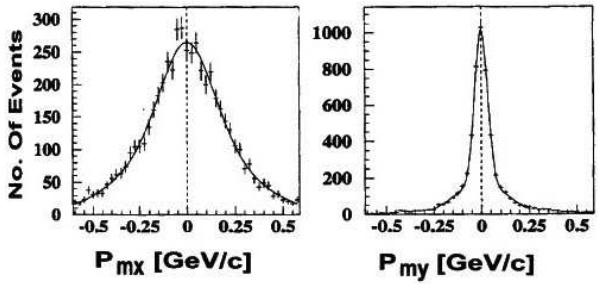}
\caption{The distribution of the transverse components of the
missing momenta, $P_{mx}$ and  $P_{my}$, for selected hydrogen
events from the $\text{CH}_2$ targets for 5.9 GeV/c. The width of the $P_{mx}$
distribution is $\sigma=0.150$ GeV/c and for the $P_{my}$ 
distribution it is $\sigma=0.035$ GeV/c.
}
\label{fig:px_py_E850}
\end{figure}

The removal of a proton with momentum, $\vec{P}_m$,
from the nucleus implies a transfer of that three momentum 
from the nucleus to the observed two-proton final state. 
A small deficit in final state energy, $\epsilon_m$, 
also can in principle be observed. 
We define the $\hat{z}$ direction to
 coincide with the incident beam direction.
The final state proton pair has transverse momentum, 
$\vec{P}_{mT}$ with the $\hat{x}$ direction 
to be in the scattering plane of 
$proton_3$ and $\hat{y}$ to be normal to this scattering plane.
By our convention, $proton_3$ has the smaller polar angle. 
So components of the nuclear momentum are:
\begin{eqnarray}
 \vec{P}_m=(P_{mx},P_{my},P_{mz}) \nonumber \\
\vec{P}_{mT}\equiv P_{mx}\hat x +P_{my}\hat y.  
\label{eqn:P_def}
\end{eqnarray}
The in-plane momentum component, $P_{mx}$, depends directly on the 
the difference of the x-components of $\vec{P}_3$ and  $\vec{P}_4$.
Hence, $P_{mx}$ has broader resolution than the out-of-plane component,
$P_{my}$, which is largely determined from the azimuthal angles.
The distribution of the $P_{mT}$
variables for the hydrogen events in $\text{CH}_2$,
is shown in Figure~\ref{fig:px_py_E850}. 

To account for the effects of longitudinal 
component of the nuclear momentum it is useful to consider the
 momentum distribution in light cone coordinates. 
Defining $E_m=\epsilon_m+m_p$, 
the light cone momentum components are derived from the nuclear momenta:
$(E_m,p_{mz}) \rightarrow (E_m \pm p_{mz})$. 
The ordinary nucleon momentum distribution
can be re-expressed as a distribution function of 
light cone components. 
The ratio  of the dimensionless light cone fraction carried by a 
single proton to that carried by the entire nucleus is $\alpha/A$. 
Specifically, we write
\begin{equation}
\alpha \equiv A \frac{(E_m-P_{mz})}{M_A} \simeq 1-\frac{P_{mz}-\epsilon_m}{m_p} 
\simeq 1-\frac{P_{mz}}{m_p}.
\label{one} 
\end{equation}
The approximate expression for $\alpha$ comes from neglecting $\epsilon_m$ and
 taking the mass of the nucleus $M_A \simeq A m_p$. 

 Analogous to the  use of pseudo-rapidity based on angles as an 
approximation to rapidity, which includes the particles' momentum, we  define 
\begin{eqnarray}
\alpha_0 & \equiv & 1-\frac{2\beta \cos{\frac{\theta_3-\theta_4}{2}}
\cos{\frac{\theta_3+\theta_4}{2}}-p_{1z}}{m_p} \nonumber \\
with \nonumber \\
\beta &\equiv & \sqrt{{(\frac{E_1+m_p}{2}})^2-m_p^2}  
\label{eqn:alphaZ}
\end{eqnarray}

\begin{figure}[htbp]
\includegraphics[width=80mm]{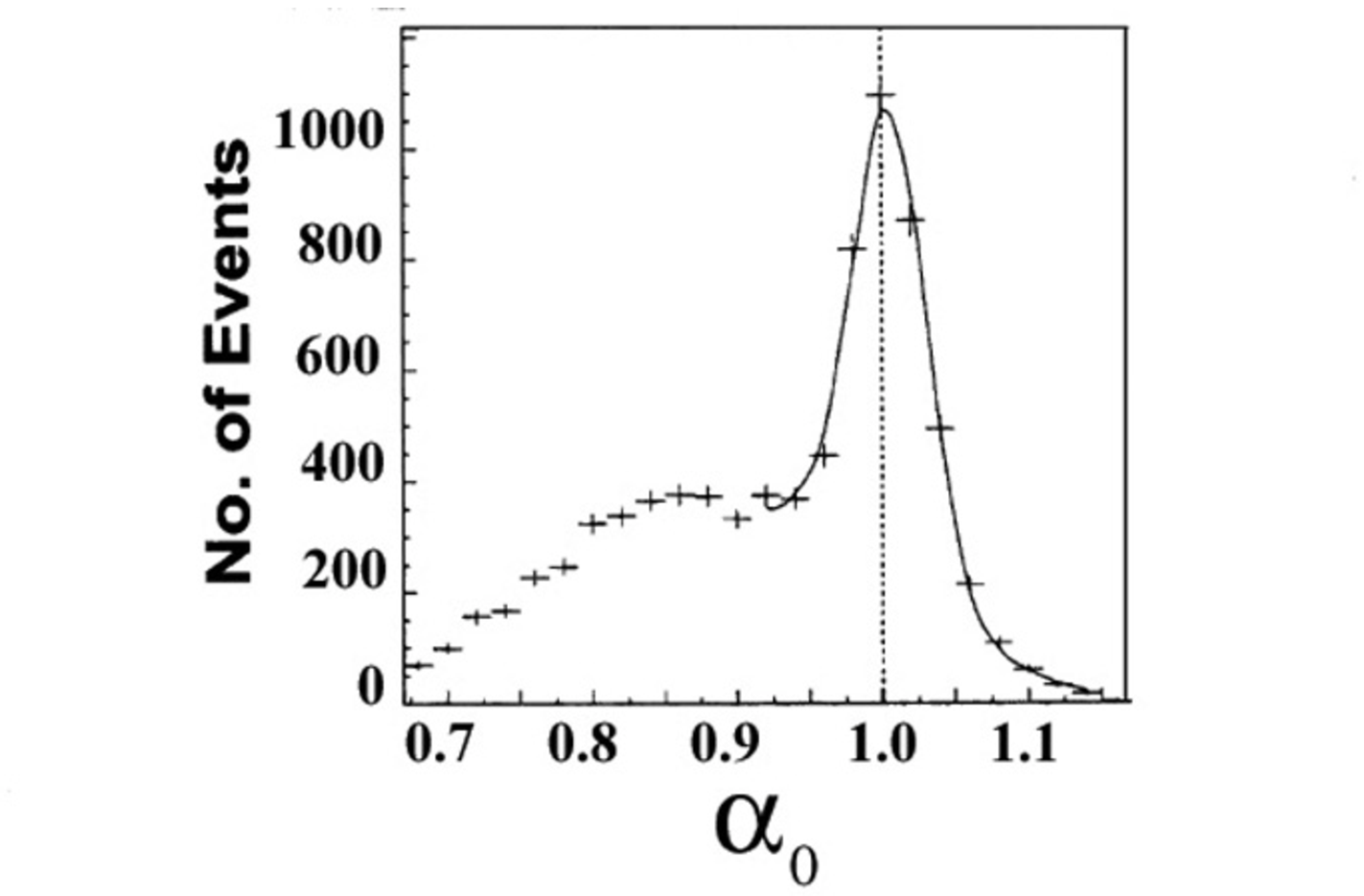}
\caption{The distribution of the longitudinal component of the
light cone momentum, $\alpha$, for selected hydrogen events
from the $\text{CH}_2$ targets. The approximation, $\alpha_0$, is substituted 
for 
$\alpha$ as described by Equation~\ref{eqn:alphaZ}.
}
\label{fig:alpha0_E850}
\end{figure}

For a given beam momentum, $\alpha_0$ is a function only of the final state
lab polar angles, $\theta_3$ and $\theta_4$, and not the final state momenta,
$P_3$ and $P_4$. 
It is obtained from the exact
expression by assuming that the final state momenta, $\beta$, are  equally 
shared
between two final state protons. For the limited range of angles of this
experiment,  $\cos{\frac{\theta_3-\theta_4}{2}}$ is
$\sim1.0$.  Since for our experiment $\alpha_0$ is better determined than
$\alpha$, we use $\alpha_0$ in our analysis.
The distribution of $\alpha_0$ for
selected $\text{CH}_2$ events at 5.9 GeV/c is shown in 
Figure~\ref{fig:alpha0_E850}.
The resolution of the  $\alpha_0$ distribution is $\sigma=0.025$, and 
calculations indicate the difference between $\alpha$ and $\alpha_0$ is less 
than 0.005 for our range of angles..

The width ($\sim200$ MeV/c) of the nuclear momentum distribution
 for the longitudinal momentum results in
approximately a 20\% spread in the 
measured $\alpha$ distribution around unity. 
Because the measured distribution is strongly influenced by the
$s^{-10}$ behavior of the  pp cross section, 
the shape of the distribution is strongly skewed toward  $\alpha < 1$. 
In the kinematic region of interest, the center of mass energy of the 
$pp \rightarrow pp$ sub-process will be nearly independent of 
$\vec{P}_{mT}$ but will depend critically upon $\alpha$.
For fixed beam energy, $E_1$, we find that $s$ (the square of the 
center of mass energy of the pp system) depends on $\alpha$ according to:
\begin{eqnarray}
s & \simeq & \alpha s_0 \nonumber \\
s_0 & = & 2 m _p E_1+2 m_p^2
\label{sEq}
\end{eqnarray}
where $s_0$ corresponds to the value of $s$ for the case of the struck proton
at rest ($\alpha=1$).
In this paper we will consider an effective incident beam momentum $P_{eff}$
calculated from an effective beam energy, $E_{eff}$:
\begin{equation}
E_{eff}=\frac{s}{2 m_p}-m_p \simeq E_1 \alpha
\label{eqn:P_eff}
\end{equation}
where the approximation reflects the relativistic limit. 
Use of the variable $P_{eff}$ has been studied in
References~\cite{Mardor:1998fz,Yael_th}.

\begin{figure}
\includegraphics[width=7.5cm]{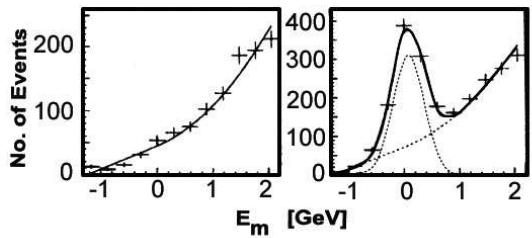}
\caption{Missing-energy, $E_m$,  distribution for $P_1$ = 5.9 GeV/c
The left hand figure is plot shows the events with one or more
extra tracks
in the straw tubes.  The right hand plot is the  $E_m$ distribution
with no extra tracks.  The dotted line represents the 
quasielastic distribution
after the interpolated background is subtracted.}
\label{fig:Emiss}
\end{figure}

We  identify the missing momentum of Equation~\ref{eqn:P_miss}
with the momentum of the nucleon in the nucleus in the
spirit of the impulse approximation.  In the longitudinal
direction this is a very good approximation. In the transverse direction,
this relation is less exact because of  elastic re-scattering.
Because the $90^\circ_{c.m.}$  $pp$ cross section
strongly depends on one longitudinal
light-cone component of the missing momentum,
we express the missing momentum in light-cone coordinates.
The coordinate system takes $\hat{z}$ as the beam direction and $\hat{y}$
normal to  the scattering plane.  In the first E850 publication regarding
nuclear transparency, the separation of signal from background was
done in the missing-energy ($E_m$)  distribution as illustrated in
Figure~\ref{fig:Emiss}
\cite{IM}.
A model for the background distribution, based on events with at least
one extra track  observed in the straw tube cylinders provided a parameterized
shape for the background subtraction.   Although this method was satisfactory 
for
the 5.9 and 7.5 GeV/c  analysis, it became less satisfactory as the incident
momentum increases and the missing energy resolution broadens.

We now describe an improved analysis procedure used for the second
publication where the background
subtraction utilizes the variation in the density of measured events per
unit four-dimensional missing-momentum space \cite{E850}. This distribution
shows a sharp quasielastic peak, and a nearly flat background.
The four-dimensional  volume element is
\begin{equation}
d\epsilon_m~d^3 \vec{P}_m \rightarrow d^2\vec{P}_{mT} d\alpha
 d(m_{M}^2)
\label{arreq}
\end{equation}
where $\vec{P}_{mT}$ is the transverse part of the missing momentum vector
as defined in Equation~\ref{eqn:P_def}, and
the longitudinal portion $\alpha$ ($\alpha_0$)
 is given by Equations~\ref{one} and \ref{eqn:alphaZ}.

\begin{figure}
\includegraphics[width=7.5cm]{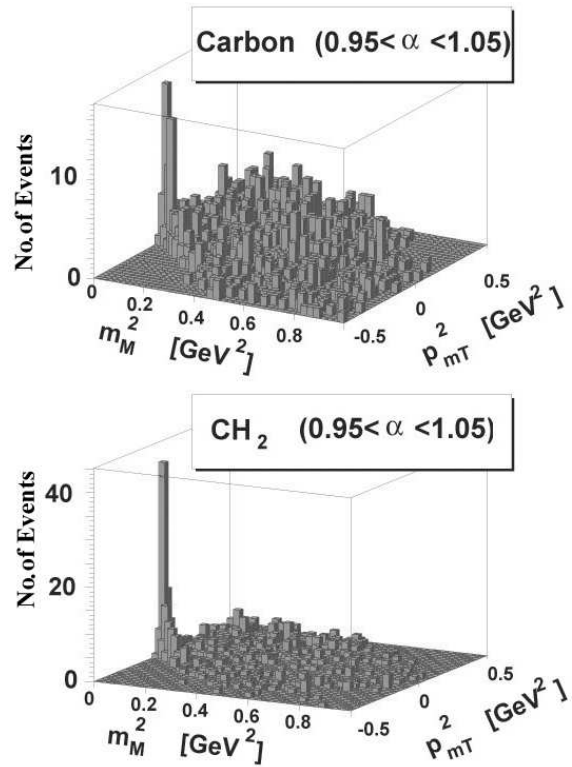}
\caption{ The distribution of $P_{mT}^2$ vs $m_M^{2}$ for  two 
tracks events observed at beam momentum of 5.9 $GeV/c$ and 
with $.95<\alpha_0<1.05$. The upper frame 
shows Carbon data and the lower frame shows CH$_2$ data.
No other kinematic cuts have been applied to these data.
}
\label{fig:mmpt}
\end{figure}

Elastic $pp$ scattering occurs at a singular point
($m_{M}^2=0$, $P_{mT}^2=0$, $\alpha=1$) in this four-dimensional phase space.
Quasielastic scattering is observed as an enhancement in a region around 
that elastic singular point in missing momentum space.
The basic idea of the extraction of elastic and quasielastic scattering
is that the four dimensional peak is distinct from the smooth background and
can be identified. In Figure \ref{fig:mmpt} the distribution of events
in $P_{mT}^2$ vs $m_M^{2}$ is presented. The only kinematic
selection criterion for events displayed in this figure
is a selection on light cone momentum for $0.95 < \alpha_0 < 1.05$. 
From Equation~\ref{arreq} we note that each
square bin in Figure \ref{fig:mmpt} represents equal four-dimensional
phase space. We assume that the background below the
 quasielastic scattering peak
is smooth.
The constant background under the quasielastic peak will be 
determined from the background level per unit phase space in the
region around the peak.

An objective of this analysis is to extract the quasielastic 
signal from background with the inclusion of events that undergo
secondary elastic scattering. In the high energy limit,
and for quasielastic scattering in
the $\alpha=1$ region, the effect of secondary elastic scattering 
is to smear the peak seen in Figure \ref{fig:mmpt} to lower $m_{M}^2$ and
larger $P_{mT}^2$. 
Studies of the data and with Monte Carlo have indicated 
 that we can  include nearly all of 
elastic rescattering in our data sample without 
accepting excessive levels of background.  

\begin{figure}
\includegraphics[width=7.5cm]{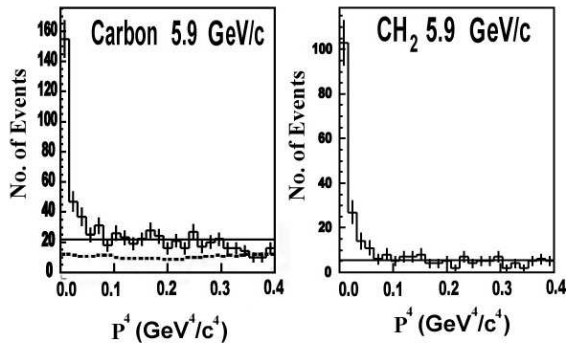}
\caption{ The distribution of {\bf P$^4$} variable for the 5.9 GeV/c
data set for Carbon and $\text{CH}_2$ targets.  The cuts of Equation~\ref{cuts}
define these selected events.  The dashed line in the Carbon distribution
shows the distribution of events, which are explicitly inelastic 
due to the presence of an extra track in the strawtubes.
} 
\label{fig:Leks}
\end{figure}

With the assumption of smooth background in the $P_{mT}^2  \times m_{M}^2$ 
plane, we extract a signal above a constant  background using the 
radial projection 
of the distribution shown in Figure \ref{fig:mmpt}.
We define the variable {\bf P$^4$} to be the sum of the squares
of the horizontal and vertical displacement from the elastic peak location,
\begin{equation}
{\bf P^4} \equiv P_{mT}^4 + m_{M}^4 .
\label{P4}
\end{equation}
The distribution of this variable corresponds to a constant four 
dimensional phase space per  $\Delta{\bf P^4}$ bin. The distribution of
the {\bf P$^4$} variable  has the quasielastic signal concentrating
near {\bf P$^4 =0$} of the distribution as shown
for the 5.9 GeV/c data in Figure~\ref{fig:Leks}  . It is quite 
natural to extend the smooth background measured in the interval
$0.15 <  {\bf P^4} < 0.35 (GeV/c)^4$ in this variable under the
quasielastic peak for ${\bf P^4} < 0.1 (GeV/c)^4$ .  
We can test this with events, which are
explicitly inelastic by selecting those which have an extra track in
the straw tube cylinders.  The dashed line plotted in 
 the 5.9 GeV/c Carbon
distribution of  Figure~\ref{fig:Leks} indicates that the background
distribution is indeed flat below the elastic peak.

\begin{figure*}
\includegraphics[width=140mm]{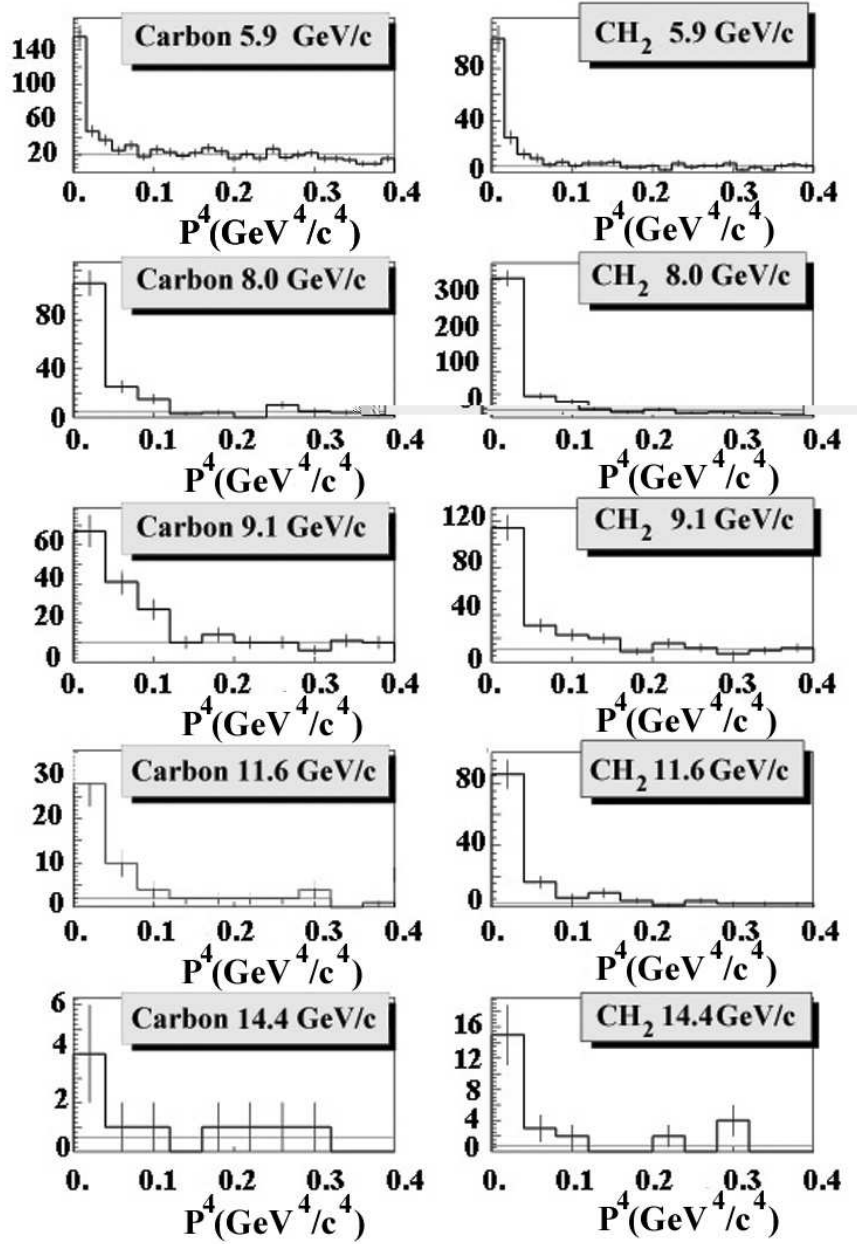}
 \caption{The {\bf P$^4$}  distribution of events as defined by the 
cuts of Equation~\ref{cuts} for Carbon and $\text{CH}_2$ targets at
each of 5 beam momenta. The horizontal lines show the level of the constant 
background
determined from  the $0.15 <  {\bf P^4} < 0.35 (GeV/c)^4$ region, and  
subtracted from under the signal peaks for  ${\bf P^4} < 0.1 (GeV/c)^4$. }
 \label{Fig10plts}
 \end{figure*}

The distributions of {\bf P$^4$} are shown in Figure~\ref{Fig10plts} for data 
collected at 5 beam momenta. The figure shows both distributions for
Carbon and $\text{CH}_2$ targets.  
For data shown in the figure, the event selection is 
defined by the selection of exactly two nearly back-to-back charged tracks in 
the spectrometer with a vertex at the appropriate target.
Kinematic cuts  applied to the events shown in Figure 
\ref{Fig10plts} are  as follows:

\begin{eqnarray}
|P_{mx}|< 0.5 \text{GeV/c,}~~~|P_{my}|< 0.3  \text{GeV/c,} \nonumber \\
~~|1-\alpha_0|<0.05 \text{,}~~~|\theta_3-\theta_4|<\Delta \theta
\label{cuts} 
\end{eqnarray}
The cut,  ($|\theta_3-\theta_4|<\Delta \theta$),
results in a range of 
c.m. scattering angles in the  pp scattering
subprocess. The modest differences in 
angular region are indicated in Table \ref{tab:E850_T} of the Appendix
for each of the 5 beam momentum data sets.
This is the same set of cuts used in 
previously published analysis \cite{IM,E850}.

The {\bf P$^4$}  distributions for both C and $\text{CH}_2$  are shown
for all 5 incident momenta  in 
Figure \ref{Fig10plts}, subject to the cuts of Equation~\ref{cuts}.
We observe that the level of background relative to signal 
remains  about 10\% even as we increase the incident momentum 
from 5.9 to 14.4 GeV/c.  This appears to indicate  that the various
inelastic background processes decrease with $s$ at least as
rapidly as the elastic or quasielastic reactions.

\subsection{ THE NUCLEAR TRANSPARENCY RATIO}
\label{subsec:TRatio}

The secondary  interactions associated with quasielastic scattering
in this energy range are
$\sim80\%$ absorptive, leading to the break up of one
or more of  the primary protons.
In this analysis elastic secondary scattering introduces minor 
perturbations in the 
trajectories of observed final state particles, but it is 
not expected to reduce the quasielastic event count if the
cuts defining quasielastic scattering are fairly 
open as given by Equation~\ref{cuts}.
We will compare our observed nuclear transparency 
to an application of the  Glauber model where
it is only the absorptive secondary interaction 
that reduces the nuclear transparency to less than unity.
For both the measurement and the Glauber calculation, one must
distinguish between the inelastic rescattering
which removes the event from the quasielastic channel
from a small-angle elastic rescattering which does not.
Independent of these difficulties, we observe
that the Glauber absorption will be independent of beam energy
in this higher energy range. 
At high energy, elastic rescattering
is known to become transverse, exchanging transverse
momentum rather than longitudinal 
momentum with spectator particles.
This will smear the transverse components of the
reconstructed final state momentum, but 
will not significantly affect the measured light cone momentum.
With this in mind, this analysis will involve
selecting data with broad cuts on transverse momentum so that elastic 
secondary interactions are
nearly all included in the event selection in agreement with Monte Carlo
simulations

The uncertainties associated with
the normalization of the nucleon density distribution is also
independent of energy. By relaxing the requirement for an
absolute normalization for nuclear transparency, we can make a 
measurement involving the ratio of cross sections for 
hydrogen and Carbon .  We first calculate the
nuclear transparency ratio to establish the energy dependence without
reference to specific assumptions about the normalization of
the nuclear momentum distribution. In the next section we will use the best
nuclear momentum distributions available to compare the absolute
normalization of nuclear transparency to the Glauber calculation.
 so that the two experiments can be directly compared.

The method of analysis for determining the C to H transparency ratio
is similar to that discussed previously \cite{IM,Israel_th}.
Signals for pp elastic and (p,2p) quasielastic 
events from runs with $\text{CH}_2$ and C targets respectively
 are extracted from the 
{\bf P$^4$} distributions  given in Figure~\ref{Fig10plts}.
 Equation~\ref{cuts} lists the kinematic cuts used for the selection.
 For this set of kinematic 
cuts, we determined $R_C$ and $R_{CH2}$, the event rates
per incident beam proton and per Carbon atom in the 
Carbon or $\text{CH}_2$ targets.
  Measurements of the
 thickness of the targets and densities allowed a determination of the 
Carbon atom densities in the two  targets. 
The yield from the two hydrogen nuclei in the 
 $\text{CH}_2$ target was $R_{CH2}-R_C$.  
 Then the experimental nuclear transparency ratio $T_{CH}$ is defined
in terms of these event rates,
\begin{equation}
T_{CH} = \frac{1}{3} \frac{R_C}{R_{CH2}-R_C}
\label{eqn:frac_CH} 
\end{equation}
The $\frac{1}{3}$ in  Equation~\ref{eqn:frac_CH} 
reflects the relative number of free protons in each $\text{CH}_2$ 
complex to number of protons in each Carbon nucleus. 


 \begin{figure}
\includegraphics[width=88mm]{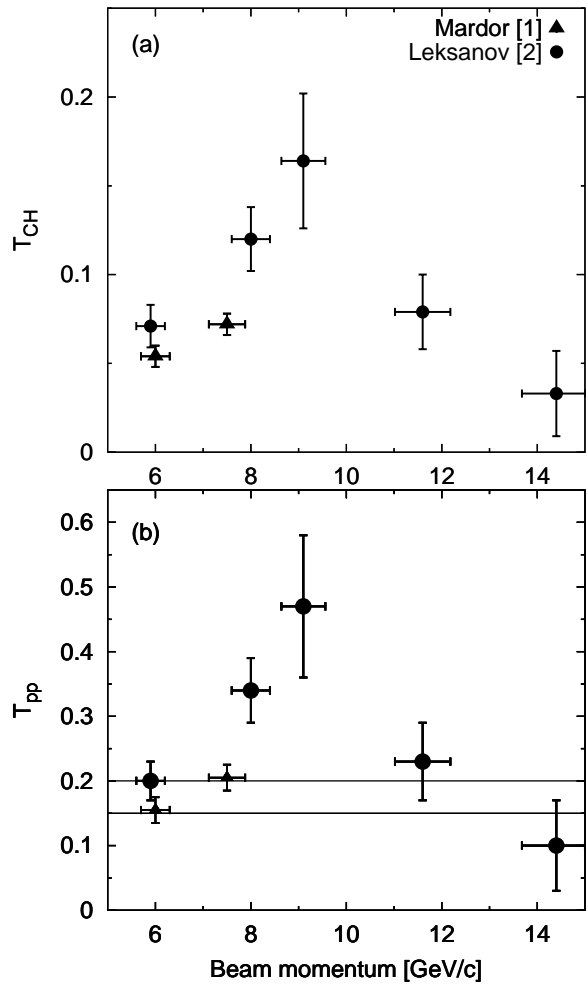}
\caption{a.(top frame) The nuclear transparency ratio $T_{CH}$ 
as a function of beam momentum.
b.(bottom frame) The nuclear transparency $T_{pp}$ as a function of the  
incident beam momentum. The events in these plots are selected using
the cuts of Equation~\ref{cuts}, and a restriction on the polar
angles as described in the text.
 The errors shown here are statistical errors, which dominate
for these measurements. 
}
\label{Fig1}
\end{figure}

There are systematic errors from our fitting and analysis procedures
which include:  The estimated error in background subtraction due to
variations in the background fitting function of 1\%.
Target misidentification due to vertex resolution of 2\%. Uncertainties
in the acceptance calculations of 1\%. The maximum beam normalization
error is less than 3\%.  Overall these combined to a 4\% error which
is small in comparison to the statistical errors.

The values for $T_{CH}$ are plotted in Figure~\ref{Fig1}a and
 listed in Table \ref{tab:E850_T} of the Appendix.
We see that there is a striking energy
dependence in the simple ratio of the rate
of Carbon quasielastic events to that of the hydrogen elastic events.

\subsection{ THE NUCLEAR MOMENTUM DISTRIBUTION}
\label{subsec:dist}

To determine the nuclear transparencies, $T_{pp}$,
we  introduce a relativistic nuclear momentum distribution.
 $n(\alpha,\vec{P}_{mT})$,   
that specifies the differential probability density per unit
four-momentum. The nuclear momentum distribution is associated with
the momentum distribution of  protons 
in the nucleus.
We discuss this function as a distribution in light cone 
momentum, where the differential element is expressed in light cone
coordinates. With Equation~\ref{six} below, we can relate the 
nuclear transparency ratio, $T_{CH}$, measured over a narrow range of 
$\alpha_1$ to  $\alpha_2$ around unity to the nuclear transparency, $T_{pp}$,
over the entire nuclear momentum distribution,

Our formulation of the nuclear momentum distribution parallels the description 
of a nonrelativistic nucleon momentum distribution by Ciofi degli Atti and 
Simula \cite{CdA}. By integrating the spectral function
 over all removal (missing) energies, 
$\text{E}_m$, they arrive at their spherically symmetric
 nucleon momentum distributions, $n(k)$; where k is the wave number.
 A calculation of the distribution, $n(k)$, requires only a knowledge of the
ground-state wave function. In our analysis, we accomplish the same task by
using a generous cut in $\text{E}_m$. As discussed in their paper, 
their nucleon momentum distributions include both the mean field,
low momentum component
of the distribution, and the high momentum component with NN correlations.
The main effect of the NN correlations is to deplete states below the
 Fermi level, and to make the states above the Fermi level partially
occupied as seen in our neutron-proton correlation measurements with the
E850 detector  \cite{Aclander:1999fd,Tang,Malki:2000gh}.  

Implicitly integrating over the missing energy, $\text{E}_m$,
 we characterize the nuclear momentum distribution,
$n(\alpha,\vec{P}_{mT})$,  over 
transverse momentum and light cone fraction $\alpha$. 
Then we introduce the integral of this nuclear momentum distribution over
the transverse coordinates:
\begin{equation}
N(\alpha)=\int \int d\vec{P}_{mT} n(\alpha,\vec{P}_{mT})
\label{N}
\end{equation}

The integrated nuclear momentum distributions
 $N(\alpha)$ can be estimated from nonrelativistic
nucleon momentum distributions of Ciofi degli Atti and 
Simula \cite{CdA}. 
 Their  parameterization for Carbon in terms of the wave number k is:
\begin{eqnarray}
n_C(k)=& \nonumber \\
&\frac{1}{4 \pi} 
(2.61 e^{-2.66 k^2}(1+3.54 k^2)+ \nonumber \\
&.426 e^{-1.60 k^2}+.0237 e^{-.22 k^2} ) 
\label{nC}
\end{eqnarray}
where the units for k and $n_C$ are $Fermi^{-1}$ and $Fermi^{-3}$ 
respectively. They also provide nucleon momentum distributions
  for $^4\text{He}$, $^{16}\text{O}$, $^{40}\text{Ca}$,
 $^{56}\text{Fe}$, and $^{208}\text{Pb}$
\cite{CdA}.  

Equation~\ref{nC} can be associated 
with the light cone distribution $N(\alpha)$ by integrating $n_C(k)$ over 
transverse momentum and by noting that near $\alpha=1$ 
as in Equation~\ref{one}. 
\begin{equation}
\alpha \simeq 1-P_{mz}/m_p.
\label{al}
\end{equation}

Knowledge of the nuclear momentum distribution represents a practical limit in
interpreting the normalization of the nuclear transparency. As we primarily 
focus on measurements around $\alpha=1$, it is the normalization of 
the momentum distribution near the origin that is most critical.
The measurement of the shape of the energy dependence of the 
nuclear transparency can be extracted with knowledge of the nuclear
momentum distribution. However for detailed comparison to the
prediction of conventional Glauber absorption, 
the quantity $N(1)$ must be known. It is 
fortunate that $N(1)$ is well constrained by a
comparison to y scaling data.

It has been pointed out
that this dimensionless normalization constant $N(1)$
is connected to  $F(y)$, the y scaling function.
The y scaling function evaluated at y=0 
is associated with a transverse integral of the
nucleon momentum distribution \cite{Frankfurt:2000ty}. The
relationship is
\begin{eqnarray}
4 \pi \int_0^\infty n_C(p) p^2 dp &= 1 \nonumber\\
2 \pi \int_0^\infty n_C(p) p dp& =\frac{1}{m_p}N(1)=F(0)
\label{YtoN}
\end{eqnarray}
The momentum distribution used here is normalized such that
$F(0)=3.3~\left( \frac{GeV}{c}\right)^{-1}$, which agrees with y scaling 
data at about the  10\% level.

We can relate the experimentally observed 
quantity $T_{CH}$ to the a convolution of the fundamental pp cross section with
a nuclear momentum distribution $n(\alpha,\vec{p}_{mT})$,
\begin{equation}
T_{CH}=T_{pp} 
\int_{\alpha_1}^{\alpha_2} d\alpha \int d^2 \vec{P}_{mT}~
n(\alpha,\vec{P}_{mT}) \frac{\frac{d \sigma}{dt}_{pp}(s(\alpha))}
{\frac{d \sigma}{dt}_{pp}(s_0)},
\label{six}
\end{equation}
where $s$ and $s_0$ are defined by Equation~\ref{sEq}.
Further noting that for fixed beam energy the ratio of pp cross sections
in Equation \ref{six} is well approximated with a function of $\alpha$ only, 
we can also write:
\begin{equation}
T_{CH}=T_{pp}
\int_{\alpha_1}^{\alpha_2} d\alpha 
N(\alpha) \frac{\frac{d \sigma}{dt}_{pp}(s(\alpha))}
{\frac{d \sigma}{dt}_{pp}(s_0)}.
\label{sixx}
\end{equation}
Finally, if the range ($\alpha_1,\alpha_2$) is restricted to a 
narrow interval around unity, we see that the relationship 
between the conventional definition of nuclear transparency $T_{pp}$ 
and the experimentally measured ratio $T_{CH}$ reduces to a simple 
proportionality,
\begin{equation}
T_{CH}\simeq T_{pp}N(1)(\alpha_2-\alpha_1).
\label{prop}
\end{equation}

Our actual determination of 
the normalization of $T_{pp}$ will
be directly obtained from Equation \ref{six} with the evaluation of the 
integral by the Monte Carlo method, including a 
weighting of the integrand
by experimental acceptance. 
The shape of the nuclear
momentum distribution, taken from work by Reference~\cite{CdA},
is used to calculate these integrals.
With the normalization fixed, a Monte Carlo program is
used to select a region of c.m. angular range where the 
geometrical acceptance is the same for elastic and quasielastic
events. Typically this corresponds to a range of $86^\circ$ 
to $90^\circ$ c.m. as given in Table~\ref{tab:E850_T} of the Appendix.

\subsection{ NUCLEAR TRANSPARENCY FOR E850}
\label{subsec:NormT}

 The evaluation of the 
integral given in Equation~\ref{six} 
 using the form the momentum distribution in Equation~\ref{nC}
yields the nuclear transparency, $T_{pp}$.  Now the measured nuclear 
transparency can be directly compared to the nuclear transparency 
calculated in the Glauber Model \cite{RG}.
 The limits of the Glauber prediction are shown as 
the two horizontal lines in Figure~\ref{Fig1}b.
The limits of the Glauber prediction and uncertainty
 were calculated using published assumptions \cite{Frankfurt:2000ty}.  
 The magnitude of the Glauber nuclear transparency is uncertain
at the level indicated but there is a general consensus 
that Glauber model predicts no significant energy dependence for 
nuclear transparency in this momentum range.
However from the pure pQCD perspective it is unclear what would
generate a scale for a peak in the nuclear transparency near 9.5 GeV/c.
The probability that the E850 result for the Carbon transparency is
consistent with the band of Glauber values is less than 0.3\%, 
and compared to a best  fit with a constant transparency of  0.24,
 the probability is less than 0.8\%.

\subsection{Deuteron Transparency}
\label{subsec:Deut_T}


For the earlier experimental run of E850, we used $\text{CD}_2$
 as well as $\text{CH}_2$
targets.  With an appropriate C subtraction we are able to
obtain a D/H ratio as given in Equation~\ref{eqn:frac_DH}, 
\begin{equation}
T_{DH} =  \frac{R_{CD2}-R_C}{R_{CH2}-R_C}.
\label{eqn:frac_DH} 
\end{equation}
 We include essentially all of the
deuteron wave function by using an expanded $\alpha_0$ interval,
$0.85 < \alpha_0 < 1.05$.  
The $T_{DH}$ transparencies for 5.9 and 7.5 GeV/c incident are 1.06 $\pm$ 0.07
and 1.10 $\pm$ 0.10 as listed in Table~\ref{tab:E850_T} of the 
Appendix. The fact that they are consistent with 1.0
 provides a further check on the 
normalization of the nuclear transparency.  Further details are to
be found in \cite{Israel_th}.


\subsection{Discussion of angular dependence}

\begin{figure}[ht]
\includegraphics[height=70mm]{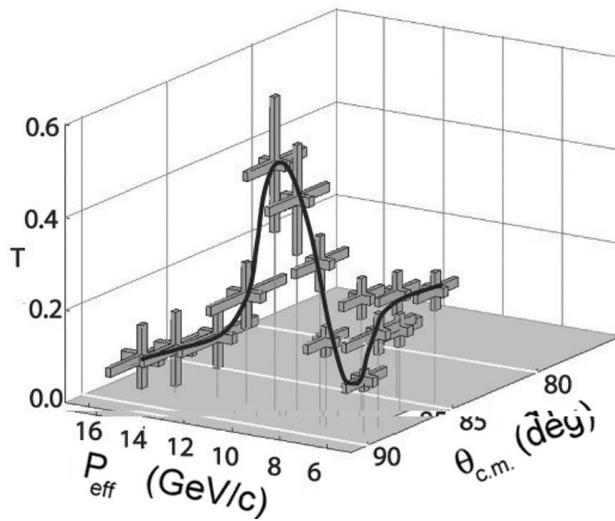}
\caption{The dependence of Carbon transparency on 
effective incident beam momentum
($P_{eff}$) and on center of mass scattering angle ($\theta_{c.m.}$).
The data are from the E850 C(p,2p) experiments.
}
\label{TvsEthcm}
\end{figure}

Figure~\ref{TvsEthcm} shows the angular dependence as well as the
momentum dependence for the Carbon transparencies from E850 as
reported in Reference~\cite{IM}.
There is a significant decrease in the
nuclear transparency at 5.9 GeV/c as $\theta_{c.m.}$ goes from
$85^\circ$ to $90^\circ$ (probability that the distribution
is flat is 0.02\%).  The poorer statistics at
7.5 GeV/c do not allow any conclusion to be drawn 
(probability that the distribution is flat is 29\%).
Measurements of the spin-spin correlation parameter, $\text{C}_{NN}$,
show rapid variation of $\text{C}_{NN}$
  with respect to the  c.m. angle near $90^\circ_{c.m.}$.
Since these spin-spin correlations are the result of 
additional scattering amplitudes,
they may also be related to changes in the nuclear transparency.
\cite{Crosbie:1981tb,BdeT,Israel_th}.

\section{Experiment E834}
This section will describe the nuclear transparency measurements with the
E834 detectors during the run of 1987.  The analysis methods employed by
 E834 will be compared to those used in E850.

\subsection{The E834 detector}
\label{subsec:E834_detect}

\begin{figure*}
\includegraphics[width=120mm]{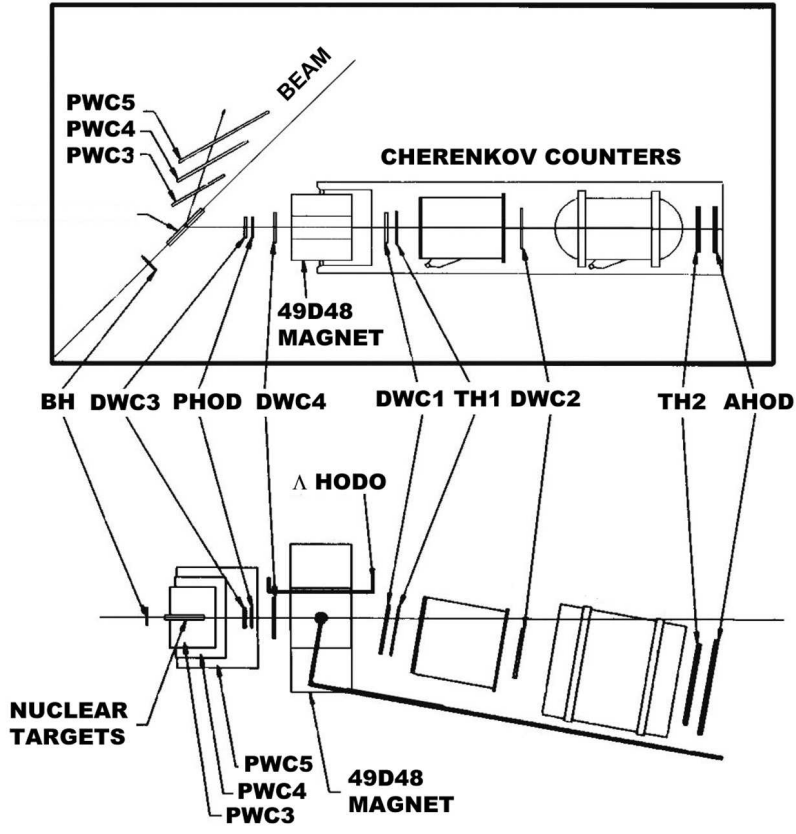}
 \caption{Schematic drawing of the top and side views of the E834 detector. 
DWC1 - DWC4 are wire drift chambers and PWC3 - 5 are proportional
wire chambers.
The two \v{C}erenkov counters detect pions and kaons in 
the spectrometer, and the
scintillation hodoscopes (BH, AHOD and TH1-2) are used for triggering. }
\label{fig:E834_detect}
\end{figure*}

The E834 detector was originally built for the measurements 
of a large number of 
different two body exclusive reactions at $\sim90^{\circ}_{c.m.}$  
with a liquid hydrogen target \cite{CW,GB,BB}.
The location of the experiment and the beam line employed was the 
same as that used later  for E850.   
As indicated in Figure~\ref{fig:E834_detect}, one long-lived,
positive particle was detected with
a high-resolution magnetic spectrometer with a resolution of
 ($\Delta{p}/p=1\%$).
Both the direction and momentum of the particle in the spectrometer were
measured with drift chambers DWC 3-4, and DWC 1-2. 
 The acceptance of spectrometer in the scattering
plane is $\Delta\theta_{lab}~\sim\pm2^0$ and $\sim5\%$ of the azimuth.
 An array of very-large-acceptance wire chambers, PWC 3-5,
measured the directions of any conjugate particles, which are 
elastically scattered or resulted from the decay of a resonance.
The acceptance of the side chambers was approximately
  $\pm30^\circ$  horizontally,
and  $\pm35^\circ$  vertically so that nearly all of the 
quasielastic distribution was
measured at one setting.  Initial fast triggering was done by the two 
scintillation hodoscopes, TH1 and TH2, with the assumption that
the interaction occurred within a small axial region of the target.
Then a more precise momentum determination for triggering is made with
the wires of DWC1 and DWC2.  The \v{C}erenkov counters in the spectrometer
arm identified pions and
kaons so that protons could be selected.
Most of the details of the wire chambers, beam and spectrometer
\v{C}erenkov counters, and the spectrometer magnet can be found in 
Reference~\cite{CW}.
When it was realized that measurements of nuclear transparency 
had an important 
relationship to exclusive reactions, this apparatus was adapted for that 
purpose \cite{E834}.

\begin{figure}
\includegraphics[width=75mm]{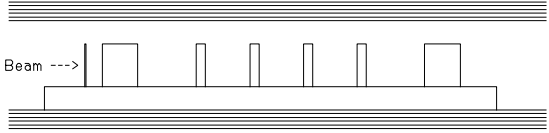}
\caption{Schematic side view of target assembly and veto counters for E834 
Experiment.  
Starting from 
the beam on the left, the elements are the final beam scintillator of 1.5 mm
thickness; a 5 cm long $\text{CH}_2$ block, four nuclear targets of either 
Li, C Al, Cu or Pb; and another 5 cm 
long $\text{CH}_2$ block.  All of these targets rest on a support of light 
aluminum 
sheet 
metal. The length of the top and bottom veto assemblies 
is 76 cm and their width is 30.5 cm.  Each veto assembly
 consists of 2 layers of lead sandwiched between 3 layers of scintillator}
\label{fig:E834_veto}
\end{figure}

The liquid hydrogen target was replaced by an array of nuclear targets 
 placed between two planes of lead-scintillator 
sandwiches to detect possible additional particles in addition to the two 
protons
from a quasielastic scattering.  As shown in Figure~\ref{fig:E834_veto}, 
there were four identical targets
 of natural isotopic abundances;  either Li, C, Al, Cu or Pb.
The number of bound protons in the four nuclear targets is
approximately 5 times the number of free protons  in the
 two 5 cm long targets of $\text{CH}_2$ on either end. 
Each of the veto planes consisted
of two layers of lead of one  radiation length sandwiched between 3, 4.8 mm 
thick scintillators.
Their size was such that 2/3 of the solid angle seen from the 
target was covered by the veto counters. Events in which charged particles or
 $\pi^0$'s produced signals in two or more layers of scintillator were 
considered
to be inelastic background.  The trigger system was set to select 
events with at least 70\% of the momentum of elastic pp events. 

  The observed distribution of vertices is shown in
 Figure~\ref{fig:E834_kinem}.
The four targets of each element
were regularly interchanged. The free hydrogen in the two  $\text{CH}_2$ blocks
 provides the normalizer for the nuclear transparency ratio.

\subsection{Kinematics for E834}
\label{subsec:Kin_E834}
The kinematic analysis of E834 proceeds along lines similar to that for E850.  
The equations for missing energy and momentum are similar to
 those used in E850 (Equations~\ref{eqn:P_miss}).

\begin{eqnarray}
 \epsilon_m = E_3 + E_4 - E_1 - m_p \nonumber \\       
 \stackrel{\rightarrow}{P}_m = \stackrel{\rightarrow}{P}_3 + 
\stackrel{\rightarrow}{P}_4 - \stackrel{\rightarrow}{P}_1.    
\label{eqn:P_miss_dup}
\end{eqnarray}

\begin{figure}
\includegraphics[width=75mm]{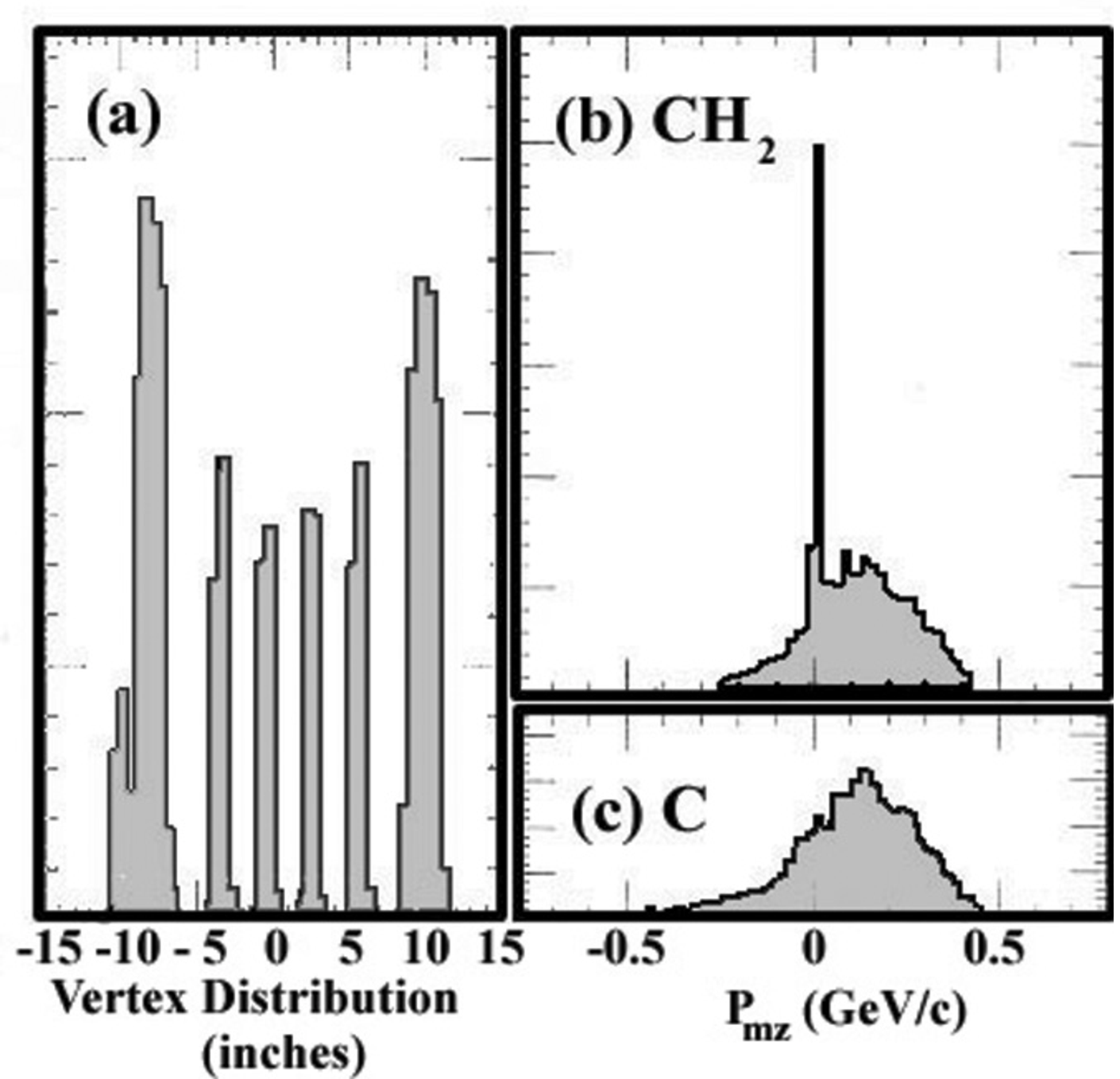}
\caption{Vertex distributions of near elastic events
 for the two $\text{CH}_2$ and four nuclear(Al)  targets~(a)
and $p_{mz}$ distributions from E834 for a $\text{CH}_2$ target~(b)
 and a Carbon target~(c). at 5.9 GeV/c.} 
\label{fig:E834_kinem}
\end{figure}

As in E850, the $\hat{z}$ component is along the incident beam direction.
 The scattering plane, containing the $\hat{x}$ component, is defined 
as the plane containing $\vec{P}_1$ and $\vec{P}_3$.  $\vec{P}_3$ is
 the particle traversing 
the magnetic spectrometer, which has no veto from the 
spectrometer \v{C}erenkovs.
The component of $\vec{P}_4$ out this plane is a measure
of $P_{my}$.  

Since there is a momentum measurement of only $\vec{P}_3$,
and not $\vec{P}_4$,
  quasielastic reactions are in principle lacking one constraint.
However, the missing energy  of the struck proton is small
compared to the momenta of the initial and final state particles. 
The components of the four momentum ${\bf P_m}$   are;
$(m_p+\epsilon_m,P_{mx},P_{my},P_{mz})$. 
We neglect $\epsilon_m$, and use energy
conservation to solve for $E_4$ in the first line of 
Equation~\ref{eqn:P_miss_dup}.
Neglecting these terms in the energy balance is a small (~0.5\%) effect in the
determination of the nuclear momenta for quasielastic events \cite{JYW}.
Figure~\ref{fig:E834_kinem}b and \ref{fig:E834_kinem}c 
show how the sharp hydrogen elastic signal is
easily extracted from the $P_{mz}$ distributions determined from the
lab polar angles of $\vec{P}_3$ and  $\vec{P}_4$.

\begin{figure}
\includegraphics[width=75mm]{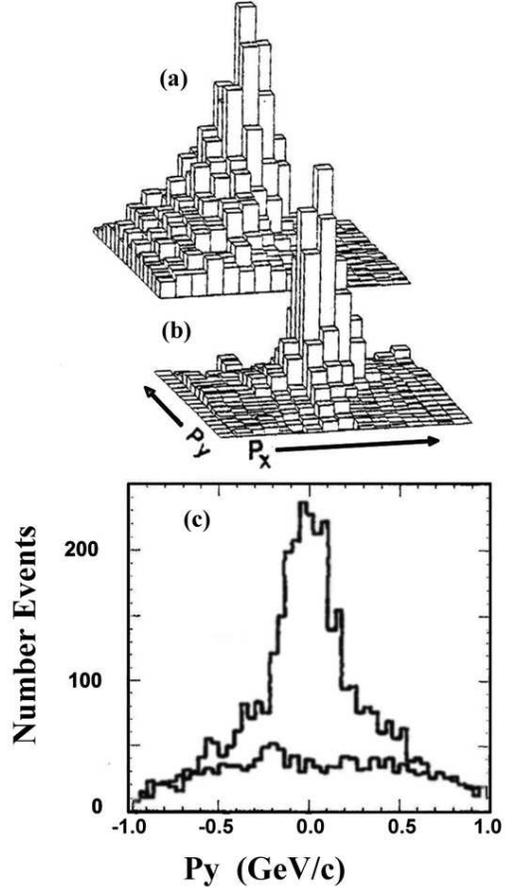}
\caption{The distribution of missing transverse momentum ($P_{mx}$ versus 
$P_{my}$) at 5.9 GeV/c
is shown for aluminum target data in the quasielastic region (a). In (b) the 
same distribution after background has been subtracted.  The projection of (a)
for events with $|P_{mx}|<0.25~GeV/c$ and $0.9<\alpha_0<1.2$ 
 is shown in (c). The lower curve represents
events with two or more hits in the veto scintillators, and the upper curve is
for events without such hits.
} 
\label{fig:pfy_bg}
\end{figure}

$P_{my}$ is mainly determined by the out-of-plane azimuthal angle,
 and is only weakly dependent on the magnitude of the 
 momentum of $P_3$ and $P_4$. 
 The difference of the magnitude of the x components of $\vec{P}_3$ 
and calculated $\vec{P}_4$ yield the error on $P_{mx}$. 
 $P_{my}$ is determined
to $\pm30$ MeV/c, while the error on  $P_{mx}$ is deduced to be 
about $\pm100$ MeV/c.
The transverse components have a negligible effect on center of mass energy
$s$, and since $\vec{P}_{my}$ is the 
better-determined component,  we plot the number of events 
versus $P_{my}$ as 
shown in Figure~\ref{fig:pfy_bg}a, b and c for Aluminum.
The `Lego plots' of missing transverse momentum ($P_{mx}$ versus 
$P_{my}$)
are shown in  Figure~\ref{fig:pfy_bg}a and \ref{fig:pfy_bg}b
 for the Aluminum target data in the quasielastic region.
 Figure~\ref{fig:pfy_bg}a presents all of the events, and
 Figure~\ref{fig:pfy_bg}b shows the distribution after background
 subtraction. 
The upper curve in Figure~\ref{fig:pfy_bg}c
displays  the events with no signals from the veto array,
and bottom curve displays those events  with veto signals in $>1$ scintillator 
planes.
The  bottom background curve in 
 Figure~\ref{fig:pfy_bg}c has been multiplied by a constant slightly
larger than unity  to match  the upper
curve for the regions $|P_{my}| > 0.50 $ GeV/c. 
 The quasielastic signal 
is the difference of these two curves.
  The events appearing this plot are selected for
$0.9<\alpha_0 <1.2$, and have $|P_{mx}| < 0.25$ GeV/c.
The events selected as quasielastic are given by the cuts in 
Equation~\ref{eqn:cuts_E834}. 

\begin{eqnarray}
|P_{mx}|< 0.25 \text{GeV/c,}~~~&|P_{my}|< 0.25  \text{GeV/c,} \nonumber \\
~~& 0.9 < \alpha_0 < 1.2.
\label{eqn:cuts_E834} 
\end{eqnarray}

The range of $0.9 < \alpha_0 < 1.2$ was selected in the E834 experiment
to provide good statistical accuracy from an interval where there is a
good signal to background ratio. Also the acceptance of quasielastic
and elastic events is nearly identical.
Since the $\text{CH}_2$ and nuclear targets were exposed to
 the same beam, there is
automatic beam normalization for the quasielastic and free pp events.

The primary systematic error associated with the E834 experiment is the 
uncertainty in the background subtraction.  The  background  
was determined from the smooth shape
fixed by the events with hits in the veto counters, and  then normalized to 
the total distribution  for  $|P_{my}| > 0.50 $ .  The background to
signal  is typically 20\% with an estimated error $\pm$5\%.   

\subsection{Nuclear Transparencies for E834}
\label{subsec:E834_T}

\begin{figure}
\includegraphics[width=90mm]{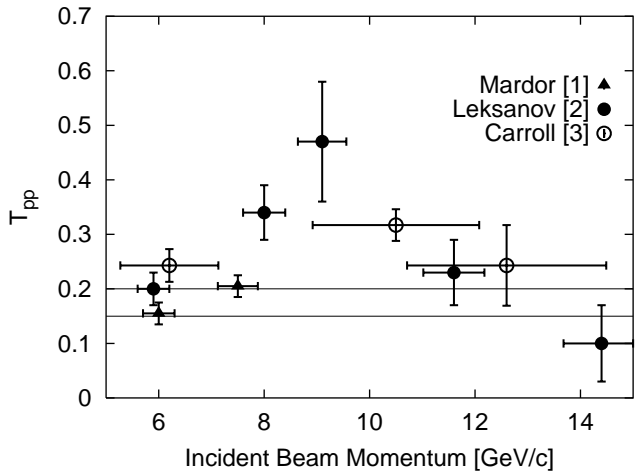}
 \caption{
Comparison of all Carbon transparency ($T_{pp}$) data from E850 and E834.
The 1988 data have been rescaled from published values 
using the momentum distribution of the form and with the normalization
described in reference \cite{CdA}.
The two horizontal lines indicate the range of values for 
$T_{pp}$ as calculated by the Glauber method \cite{RG}. 
The horizontal error bars represent the total spread in
effective momentum resulting from the accepted $\alpha_0$
range. 
 }
 \label{AllT}
 \end{figure}

Figure~\ref{AllT} shows the comparison of the Carbon transparency 
measurement of E850
to that reported in our E834 paper. The E834 analysis used a form
for the nucleon nuclear momentum distributions, which consisted of
35\% hard sphere with radius of 0.22 GeV/c for Carbon and 65\% Gaussian with
standard deviation of 0.25 GeV/c fitted to our experimental
results  \cite{SpecFn}. The form of the function, $n_G(p)$, is given below:
\begin{equation}
n_G(p)=\frac{f_f}{[(4/3)\pi{p_f^3}]}\Theta(p-p_f)+
\frac{(1-f_f)}{[2\pi{p_G^2}]^{3/2}}{exp[-\frac{p^2}{2{p_G^2}}]}\text{,}
\label{eqn:Gauss_fn}
\end{equation}
where $f_f$ is the fraction of the distribution in the Fermi gas distribution; 
$p_f$ is the radius of the Fermi sphere (0.22 GeV/c for Carbon); and
$p_G$ is the radius of the spherical Gaussian (0.25 GeV/c). The step function,
$\Theta(p-p_f)$, is 1 inside the Fermi sphere and 0 outside. 
The relative proportion of hard sphere to Gaussian
remains the same, but the radius of the hard sphere is varied from 0.170 to
0.260 GeV/c for Li to Pb nuclei.  For this present publication
 the E834 data  are made consistent with the E850 nuclear transparencies
by multiplying by the ratio of integrated nuclear momentum distributions
 given in Equation~\ref{eqn:ratio_E850_E834}.
\begin{equation}
\frac{\int_{pp}}{\int_{E834}}=
\frac{\int_{\alpha_1}^{\alpha_2} d\alpha \int d^2 \vec{P}_{mT}~
n_{CS}(\alpha,\vec{P}_{mT}) \frac{\frac{d \sigma}{dt}_{pp}(s(\alpha))}
{\frac{d \sigma}{dt}_{pp}(s_0)}}
{\int_{\alpha_1}^{\alpha_2} d\alpha \int d^2 \vec{P}_{mT}~
n_G(\alpha,\vec{P}_{mT}) \frac{\frac{d \sigma}{dt}_{pp}(s(\alpha))}
{\frac{d \sigma}{dt}_{pp}(s_0)}}
\label{eqn:ratio_E850_E834}
\end{equation}
Where $n_G$ is the spherical 
Gaussian distribution used in E834 as given in 
Equation~\ref{eqn:Gauss_fn}. 
 $n_{CS}$ are the nuclear momentum distributions derived from
 Ciofi degli Atti 
and Simula of the forms given in Equation~\ref{nC}. 
The E834 transparencies from  events in the range of $0.9 < \alpha < 1.2$. 
 listed for Li, C, Al, Cu and Pb are included in Table~\ref{tab:E834_nuc}
of the Appendix.

\section{Energy Dependence of the Transparencies from E834 and E850 }

\subsection{Systematic Errors Due to Theoretical Uncertainties}
\label{subsec:Th_Un}
Before we present the theoretical interpretations of our combined results,
we will discuss some of the uncertainties connected with the theoretical
treatment of elastic scattering from protons bound in nuclei. We have attempted
 to minimize their effects through our methods of analysis.

As noted earlier, the extraction of nuclear transparency depends on the 
assumption that
the scattering process can be factorized into a product of two functions; 
the free pp scattering at an appropriate center of mass energy, $s$, 
and a nuclear momentum distribution,
n($\alpha$,$\vec{P}_{mT}$).  This is expected to be a good 
approximation at high incident momenta
and large momentum transfers where the impulse approximation is valid.

In our first 1988 publication we 
quoted an overall error due to possible off-shell and
momentum distribution uncertainties of $\pm25\%$ \cite{E834}.  Since then the 
knowledge of nuclear momentum distributions has been considerably improved.
For the combined experiments we have settled on the recent parameterization
by  Ciofi degli Atti
 and Simula of  the nucleon momentum distributions \cite{CdA}.
  Their nucleon momentum distributions
 are based on a careful study of electron scattering experiments. 
The nuclear momentum distribution
 has been measured to large values of the nuclear momentum by studying the
 enhanced contribution from values of $\alpha$ less than 1.0.  We found 
equivalent  distributions
 in both the transverse and longitudinal directions \cite{Mardor:1998fz}, 
\cite{Yaron}.
 The high degree of two-body correlations above the Carbon Fermi 
level of 220 MeV/c is dramatically verified in our measurement, 
which correlated neutron momentum with the
 proton nuclear momentum determined in C(p,2p) reactions with the E850 detector
 \cite{Aclander:1999fd}, \cite{Tang}.

 How to deal with the off-shell nature of the pp scattering 
in the nucleus is a difficult question. Fortunately, the energies of the 
experiment are large compared to the
 nuclear binding energies.  However, the $s^{-10}$ energy dependence of the pp 
cross section magnifies even relatively small effects in the longitudinal 
direction.  The effect of the struck proton being off-shell is liable to 
suppress the 3-quark Fock state as gluons
 are exchanged to bring it on-shell.  The effect of the binding on the 
effective $s$ of the interaction, also needs to be considered. 

 We estimate a systematic error of  $\pm15\%$ for these theoretical 
uncertainties for this present publication.

%


 \subsection{Combined Nuclear Transparency Data}
\label{subsec:Comb_T}

After unifying the normalization of the E834 and E850 transparencies,
 we can plot the Carbon
transparency results for both experiments as shown in Figure~\ref{AllT}. 
If we interpolate the E850 results to compare to the E834 transparencies
at the same incident momentum,
then we find very good agreement as to their magnitudes despite the 
different analysis methods and $\alpha$ ranges.
  The ratios of $\frac{T_{pp}(E850)}{T_{pp}(E834)}$
at 6.2, 10.5 and 12.6 GeV/c are: 0.91$\pm$0.12, 1.06$\pm$0.20,
and 0.76$\pm$0.30.
 
The most striking aspect of the these two experiments
 is the confirmation of 
the peak in nuclear transparency at about 9.5 GeV/c incident momentum.  
Neither the Glauber model nor the naive prediction 
of the Color Transparency model can explain the data.
The range of the conventional Glauber calculation is indicated on
the figure by the two horizontal lines  at
 0.15 and 0.20 \cite{Frankfurt:2000ty}.

\begin{figure*}
\includegraphics[width=110mm]{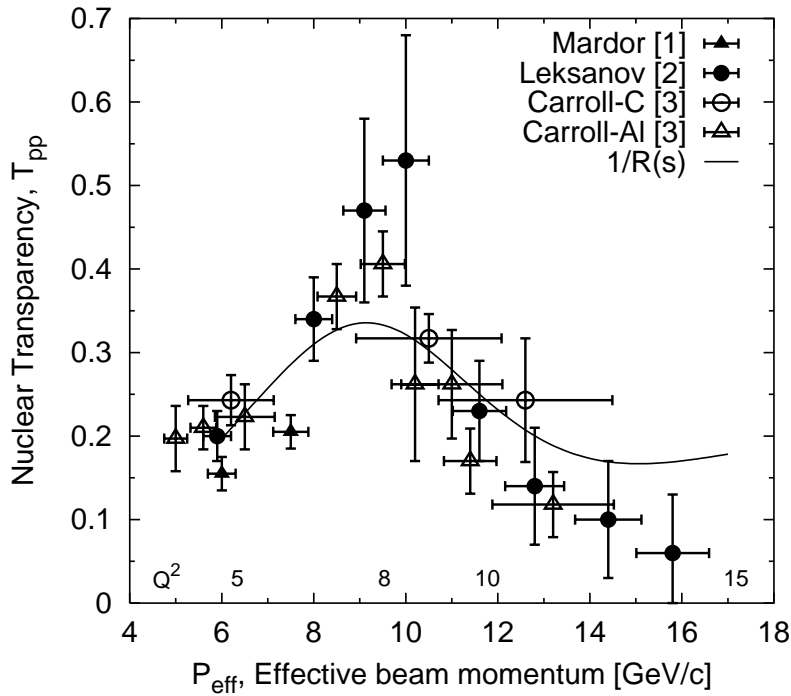}
\caption{
The $T_{pp}$ values for Carbon and the aluminum (scaled by $(27/12)^{1/3}$)
are plotted versus their $P_{eff}$ values. For a single incident
beam momentum, $P_1$, a range of  $P_{eff}$ values is obtained
by using Equation~\ref{eqn:P_eff}. This allows us to place more
points on the nuclear transparency curve and extend the range of momenta.
 The  curved line is the inverse
of R(s) defined by Equation~\ref{eqn:Rs}, and adjusted 
with an  amplitude for a 
best fit to the magnitudes of the  measured transparencies. 
The horizontal error bars represent the total spread in
effective momentum resulting from the accepted $\alpha_0$ range. 
A  $Q^2$ $[(GeV/c)^2]$ scale  is included at the bottom of the figure. }
\label{T_peff}
\end{figure*}

To further emphasize the emerging pattern, 
we add the aluminum data  measured in E834 to all the
Carbon results. For consistency, we again must make the correction
for differences in nuclear momentum distribution used here and in 
the E834 analysis. In Table \ref{tab:E834_Al} of the Appendix,
 the E834 data is corrected first
with a ratio of the new to old nuclear momentum distribution integrals. 
Then from the 1988 analysis of the E834 data
it was determined that at fixed energy, 
the dependence of the  nuclear transparency was compatible with the
inverse of the nuclear radius,  $A^{-1/3}$.
So  the statistically more  precise 
Aluminum data can be compared to the Carbon transparencies
by multiplying by this expected A dependence, $(27/12)^{1/3}$
as shown in Figure~\ref{T_peff}.

When we include the normalization and shape  of the nuclear
momentum distribution, we obtain not only transparencies,
$T_{pp}$, at $\alpha=1$ for the nominal incident momentum, but 
we can also interpret data for $\alpha \neq 1.0$.
We note from Equation \ref{eqn:P_eff} that 
with a fixed incident proton momentum, 
it is possible to study quasielastic scattering at 
an extended range of center of mass energies, corresponding 
to effective beam momentum, $P_{eff}$, above 
and below the nominal value.

In  Figure~\ref{T_peff}
we fully exploit our knowledge of the energy dependence of the pp cross 
section and the nuclear momentum distribution to extract transparencies from the 
9, 11.6 and 14.4 GeV runs at larger values of  $\alpha$
 ($1.05<\alpha<1.15$). These
points are plotted at their corresponding $P_{eff}$. We see that they fill in
to form a smooth curve extending to higher energy. 
In E834, $P_{eff}$ was calculated  for 3 bins in $\alpha$ 
(0.8-0.9, 0.9-1.0, and 1.0-1.2)
as given in Table~\ref{tab:E834_Al} of the Appendix.
  These points also fall on a smooth curve between
the points corresponding to $\alpha=1.0$. Figure~\ref{T_peff} 
demonstrates that there
is a peak in the nuclear transparency at 9.5  GeV/c nearly independent of 
whether the 
combined data set, or the separate E850 and E834 sets are used.  
Beyond 9.5 GeV/c the nuclear transparency returns to the Glauber level 
or below at 12 GeV/c and higher momentum.

\subsection{Discussion of Energy Dependence}
\label{subsec:E_dep}

The initial rise in the nuclear transparency
between 5.9 and 9.5 GeV/c was thought to be a 
manifestation of Color Transparency, namely the expansion of 
a very small configuration of valence quarks over distances
comparable to the nuclear radius.  
  In expansion models, the high $P_T$ interaction is presumed to select 
nearly point like configurations (plc's) of valence quarks in the 
 interacting protons \cite{MB}.   These plc's proceed to 
expand as they recede from the point of interaction.
  The rate of expansion has been described in 
both partonic and hadronic representations \cite{FLFS},\cite{JM}. 
Farrar, Liu, Frankfurt, and Strikman suggested a convenient  
expansion parameterization given by
 Equation~\ref{eq:S} \cite{FLFS}:  
\begin{widetext}
\begin{equation}
\sigma_{int}^{eff}(z,Q^2)=
\sigma_{int}\left(\left[\left(\frac{z}{l_h}\right)^{\tau}+
\left(\frac{r_{t}(Q^{2})^{2}}{r_{t}^{2}}\right)\left(1-
\left(\frac{z}{l_h}\right)
^{\tau}\right)\right]\theta\left(l_h-z\right)+\theta\left(z-l_h\right)\right)
\label{eq:S}
\end{equation}
\end{widetext}
where $\sigma_{int}$ is the free pN interaction cross section,
 $l_h$ is the expansion distance of the protons,
 and z is the distance from the interaction point.
 $\sigma_{int}^{eff}(z,Q^2)$  expands linearly or 
quadratically from its initial size depending on 
the value of $\tau$, and then assumes the
free space value,  $\sigma_{int}$, when $z=l_h$.  
The actual value of $\sigma_{int}$
used in the fitting procedure is less than the 
free pN total cross section, $\sigma_{tot}$(pN) $\sim40mb$,
 because nearly all ($\sim90\%$) of the
elastic cross section is within the acceptance of our detector.
The exponent $\tau$ allows 
for two suggested pictures of expansion.  For $\tau$ = 1,
 the expansion corresponds to the quantum diffusion 
picture \cite{FLFS}. For this picture, $l_h$ = $2p_f$/$\Delta$($M^2$)
where $p_f$ is the momentum of a  proton traveling 
through the nucleus \cite{FLFS}. At distances comparable to 
nuclear sizes, the effective cross sections should revert 
to their free space values. The authors of  \cite{FLFS} indicate
the values of   $\Delta$($M^2$)   between 0.5 and 1.1 $GeV^2$  
are acceptable with  $\Delta$($M^2$)=0.7
being favored.  This range of 
 $\Delta$($M^2$)  corresponds to 
values of  $l_h$ =  0.36$p_f$ to  0.78$p_f$ fm.  For a 
momentum of 5.9 GeV/c the expansion distance will be 
between 2.1 and 4.7 fm.  .  

The case of $\tau=2$ is considered to be the 
`naive quark expansion' scenario in which the light
quarks fly apart at a maximum rate and the distance 
is determined by the Lorentz boost to the hadrons.
In this case $l_h=\sim{E/m_h}$ where $m_h$ is 
the mass of the hadron
involved \cite{FLFS}. For protons at 5.9 GeV/c, the 
expected expansion distance is $\sim7.3$ fm.
   The quantity  $<r_{t}(Q^{2})^{2}>/<r_{t}^{2}>$ 
represents the  fraction of $\sigma_{int}$ at the time of interaction.  
This quantity is approximated by $\sim{1/Q^2}$, corresponding 
to 0.21 at 5.9 Gev/c and 
falling  with an increase of incident momentum \cite{FLFS}. 

Since we are dealing with hadrons, Jennings and Miller reasonably suggested
that a hadronic representation of the interacting protons
 should be considered \cite{JM}. In this picture,  
the expansion distance at 5.9 GeV/c should correspond to 
$\sim{0.9}$ fm for linear expansion, and 
 $\sim{2.4}$ fm for the quadratic 
case depending on the form of their intermediate state
 $g(M^{2}_{x})$ \cite{JM}. The shape of the 
expansion is approximated by that given in Equation~\ref{eq:S}.

However, the drop in the nuclear transparency above 9.5 GeV/c requires
additional mechanisms. 
We will discuss two possible explanations 
that have been introduced for the energy dependence. 
First, Ralston and Pire 
\cite{Ralston:1988rb} 
suggested that the structure of the nuclear transparency may arise from an 
interference between two distinct amplitudes 
that contribute to pp elastic 
scattering. One amplitude is the hard  amplitude that we 
associate with quarks at small transverse spatial separation
for dimensional scaling, which
should dominate the high energy cross section.
The other amplitude is a soft component, essentially a remnant of 
the higher order radiative (Sudakov) process that strongly
attenuates the large distance part of the proton wave function.

The dependence of the pp elastic cross section near $90^\circ_{c.m.}$ 
degrees varies nearly as a power of $s$:
\begin{equation}
\frac{d\sigma}{dt_{pp}}(\theta=90^\circ_{c.m.}) = R(s)s^{-10}.
\label{eqn:Rs}
\end{equation} 
Compared to the pp cross section, which varies over 5 orders of 
magnitude, the variation in R(s) is only about a factor
of two in the region of $s$ spanned by these
experiments \cite{Hendry:1974ex},\cite{Ralston:1988rb}. 
The dependence of the cross section on $\alpha$ will reflect both the
nuclear momentum distribution(Equation~\ref{nC}) and the cross 
section dependence of (Equation~\ref{eqn:Rs}).
The short distance part is predicted to have an energy dependency that scales 
as ($s^{-10}$). In the picture of Ralston and Pire  the long-ranged portion of 
the amplitude is attenuated by the nuclear matter,
 and the interference largely disappears for the numerator of
Figure~\ref{fig:Tr_illust} but not in the free pp scattering of the
denominator \cite{Ralston:1988rb}. 
Therefore these  authors predicted that this picture would naturally
lead to an energy dependence of nuclear transparency, which could vary like
$R(s)^{-1}$ where $R$ is defined in Equation \ref{eqn:Rs}.  
The curve $R^{-1}(s)$) is shown on Figure \ref{T_peff}
arbitrarily normalized to the approximate scale of the nuclear transparency.
This model suggests that the nuclear transparency should rise again at 
$\sim20$ GeV/c.

Another possible perspective on the energy dependence 
is suggested by Brodsky and deTeramond \cite{Brodsky:1988xw}. They observed
that the excess in the pp elastic cross section above the scaled cross section
associated with a peak in $R(s)$ at beam momenta above 9.5 GeV/c
could correspond to a resonance or threshold for a new scale of
physics. They suggested that one candidate involved the 
charmed quark mass scale corresponding to $\sim12$ GeV/c incident protons,
but other exotic QCD multi-quark states could also be considered.
They connected this resonance model with the anomalous spin dependence
of the pp elastic cross section at these 
energies \cite{Crosbie:1981tb}. In their
model, the increase in nuclear transparency from 5.9 to 9.5 GeV/c is associated
with ordinary color transparency but
the rapid reduction of nuclear transparency back to
Glauber levels would signal a contribution from a new
mass scale or new dynamic QCD scale. The connection to the polarization
dependence of these cross sections indicates that the new scale
behaves like a spin 1 resonance.

We note that our measurement of $T_{pp}$ in the peak 
region $\sim9.5$ GeV/c is approaching 50\%, and is more 
than twice the Glauber level. 
At the peak, we find that the A(p,2p) Carbon
transparency is almost as large as
the (e,e'p) nuclear transparency as described in Section~\ref{sec:other_exp}
later. This comparison is surprising and interesting 
even without reference to a detailed Glauber calculation.
 
In conclusion, we note that both the Ralston - Pire model and
the Brodsky - deTeramond picture can probably accommodate our measurement.
The Brodsky - deTeramond model, however, would predict a dramatic dependence
of nuclear transparency upon the initial state spins of the protons.
 We believe that a double spin measurement the nuclear transparency of light
polarized nuclei may be the best way to distinguish these two 
models \cite{AGS_2000}.  A measurement of the A-dependence at 12 GeV/c and
above as described in the next section would be valuable. In the 
Ralston-Pire picture, the absorption cross section would continue
to decrease from previous values  \cite{Ralston:1988rb}.

\section{A Dependence from E834}  
\label{sec:A_dep}

\begin{figure}
\includegraphics[width=75mm]{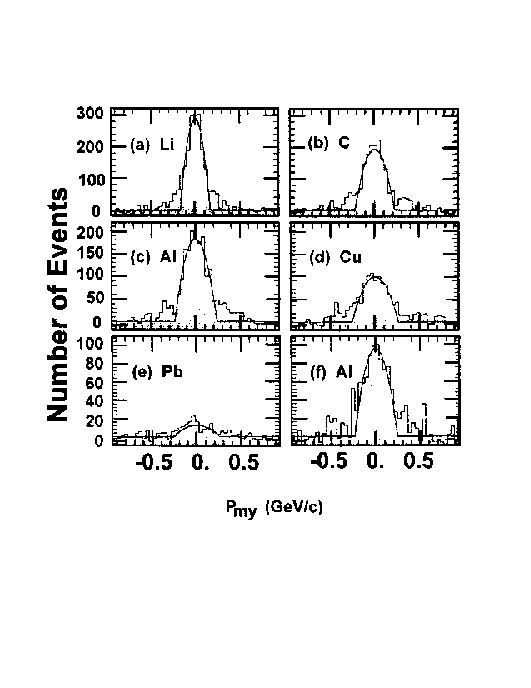}
\caption{The missing $P_{my}$ distributions for the 5 elements
used in  E834 
after corrections for acceptance and for background subtraction. 
Distributions for (a) to (e) are at 5.9 GeV/c, and the
distribution (f) is at 10 GeV/c. The form
of the model of Moniz, et al is superimposed over each distribution 
 \cite{Moniz}. }
\label{fig:Li_Pb_distr}
\end{figure}

\begin{figure}
\includegraphics[width=65mm]{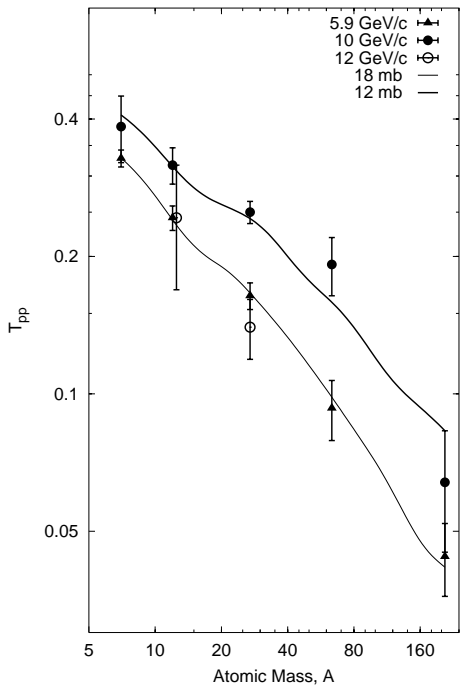}
 \caption{Nuclear transparency vs Atomic Mass A for the A(p,2p) measurements 
from E834
 for incident momenta of 5.9, 10 and 12 GeV/c as indicated on the figure.  The 
error bars reflect the
statistical uncertainties and a 10\% target-to-target systematic error.  The 
solid curves represent the fits with constant effective cross sections 
to the five nuclei at 5.9 and 10 GeV/c as described in the text. 
}
\label{fig:A_dep}
\end{figure}

As well as observing the energy dependence of the Carbon transparency, we also 
determined the nuclear transparency of Li, Al, Cu, and Pb at incident 
momenta of 5.9 and 10 GeV/c.  Transparencies for Carbon and Aluminum are also
determined at 12 GeV/c, but the Carbon transparency has a relatively large 
error since this transparency was determined from the $\text{CH}_2$ 
targets rather than a pure Carbon target.. 
The wide range of radii from Lithium (2.2 fm) 
to Lead (6.5 fm) enables us to determine the
attenuation of protons in nuclear matter, and to determine in an independent 
manner that the effective absorption cross section 
is less than that used in Glauber calculations.
The numbers of quasielastic events for all the 5 nuclei are extracted in the
same manner as for the Aluminum targets in Subsection~\ref{subsec:Kin_E834}.
The cores of the $P_{my}$ distributions are compatible with known electron
distributions as seen in Figure~\ref{fig:Li_Pb_distr} \cite{SpecFn}.
We list in  Table~\ref{tab:E834_nuc} and plot in Figure~\ref{fig:A_dep}
the transparencies for these 5 nuclei for the range of $0.9 < \alpha < 1.2$.
These transparencies have been normalized by multiplying by the ratios 
of the integrals
using Ciofi degli Atti and Simula nucleon momentum distributions to 
the integrals using 
the nuclear momentum distributions of the original 1988 analysis as given by
Equation~\ref{eqn:Gauss_fn}. 
For Li, Al and Cu an interpolation
between the published nucleon momentum distributions
 is applied to the integrals.
For Li we use $^4\text{He}$ and $^{16}\text{O}$; for Al
 it is between $^{16}\text{O}$
 and $^{40}\text{Ca}$;
and for Cu it is between $^{56}\text{Fe}$
 and $^{208}\text{Pb}$ \cite{CdA}.

As can be seen in Figure~\ref{fig:A_dep}, the 10 GeV/c transparencies
of all 5 nuclei  are consistently higher than those at
5.9 and 12 GeV/c, in agreement with the energy dependence 
observed with the Carbon and Aluminum data.
The solid lines passing through the 5.9 and 10 GeV/c data points 
represent fits with a constant 
effective cross section of $17.9^{+2.7}_{-1.5}$ mb at 5.9 GeV/c and 
$12.3^{+2.6}_{-2.6}$ for
10 GeV/c, and a floating normalization \cite{p_exp}.  
These values are consistent with the qualitative analysis performed by 
Heppelmann \cite{SH}, and a detailed analysis 
by Jain and Ralston \cite{pJ}.  Further analysis by Carroll found that
most expansion models are not compatible with our measured transparencies
 \cite{p_exp} .

This observation of the energy variation of the attenuation is an independent 
indication that the absorption of protons for large $Q^2$ quasielastic events
 is less than that predicted by free nucleon-nucleon scattering.
The errors on the transparencies of the two adjacent nuclei at 12 GeV/c do not 
allow us to measure the attenuation for this momentum.

\section{$(\pi^{+},\pi^{+}p)$ Transparencies}

Both E850 and E834 are were performed in secondary hadron beams, and used 
differential \v{C}erenkov counters for beam particle identification.  The  
$(\pi^{+},\pi^{+}p)$ $90^{o}$ differential cross sections are  10 to
25 times smaller than the (p,2p) cross sections.  However, we were 
able to make initial
measurements of the ratios of the A$(\pi^{+},\pi^{+}p)$ to A(p,2p) 
transparencies for all 5 targets at 5.9 GeV/c,
and for the Al target at 
10 GeV/c.  We multiply by the A(p,2p) transparencies
to obtain the transparencies plotted in Figure~\ref{fig:pi-p_transp};
these are also tabulated and presented
in Table~\ref{tab:pi-p_Tr}  of the Appendix \cite{JYW}.
The ratios are measured
on the slightly larger interval, $0.8 < \alpha_0 < 1.2 $, 
but the effect is negligible compared to the statistical errors.
Due to the large statistical errors, and the measurement of a single
 transparency at 10 GeV/c, attempts to determine effective cross sections from 
these  A$(\pi^{+},\pi^{+}p)$ transparencies were not useful as they
were for the A(p,2p) transparencies in the previous section.
  We only note that
the  A$(\pi^{+},\pi^{+}p)$ transparencies are $\sim1.5\times$
the corresponding A(p,2p) transparencies.
More details, including the identification of the $\pi^+$ in the 
beam by the differential counters, can be found in the thesis by
J-Y Wu \cite{JYW}.

\begin{figure}[hbt]
\includegraphics[width=65mm]{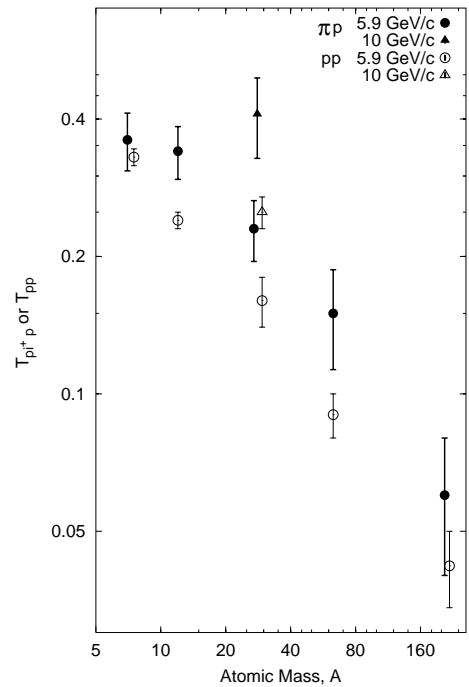}
\caption{$T_{\pi^+p}$ and $T_{pp}$ transparencies 
for Li, C, Al, Cu and Pb at
5.9 GeV/c, and Al for 10 GeV/c.  The $T_{\pi^+p}$ values 
(solid symbols) are consistently larger than those of
 $T_{pp}$ (open symbols).
} 
\label{fig:pi-p_transp}
\end{figure}


\section{Comparison to other Experiments}
\label{sec:other_exp}
The A(p,2p) experiment of Tanihata, {\it et al.} had
 1.46 GeV/c protons incident
on a set of 5 different nuclei (ranging from C to Pb in size); their results 
indicate a mean free path of $\sim2.4$ fm \cite{ITAN}. With a nuclear
density of $\frac{1}{6}$ $fm^{-3}$, this corresponds to $\sigma=25mb$.
The $\sigma_{tot}$ for the two outgoing protons of 0.95 GeV/c is $\sim25 mb$ 
\cite{sig_tot}.
We therefore conclude that the value of the nuclear transparency, below 
the minimum momentum studied in our experiments, 
is close to the Glauber limit. 

The other major investigations of nuclear transparency
have been with quasielastic scattering in the A(e,e'p) reaction
at SLAC \cite{SLAC1,SLAC2}, and at Jefferson Lab \cite{Abbott,Jlab}.
In these experiments, the nuclear transparency, $T_{ep}$,
 as a function of $Q^2$ is
\begin{equation}
T_{ep}(Q^2) = \frac{\int_{V}d^3P_mdE_mY_{exp}(E_m,\vec{P}_m)}
{\int_{V}d^3P_mdE_mY_{PWIA}(E_m,\vec{P}_m)}
\label{eqn:Tep}
\end{equation}   
where $Y_{exp}$ is the experimental yield of quasielastic scatters, and
$Y_{PWIA}$ is the calculated yield in the partial wave impulse approximation.
These studies use a prescription for the off-shell $\sigma_{ep}$. 
However, inserting the on-shell form changes $T_{ep}(Q^2)$ by less than 1\%. 

\begin{figure}[hbt]
\includegraphics[width=75mm]{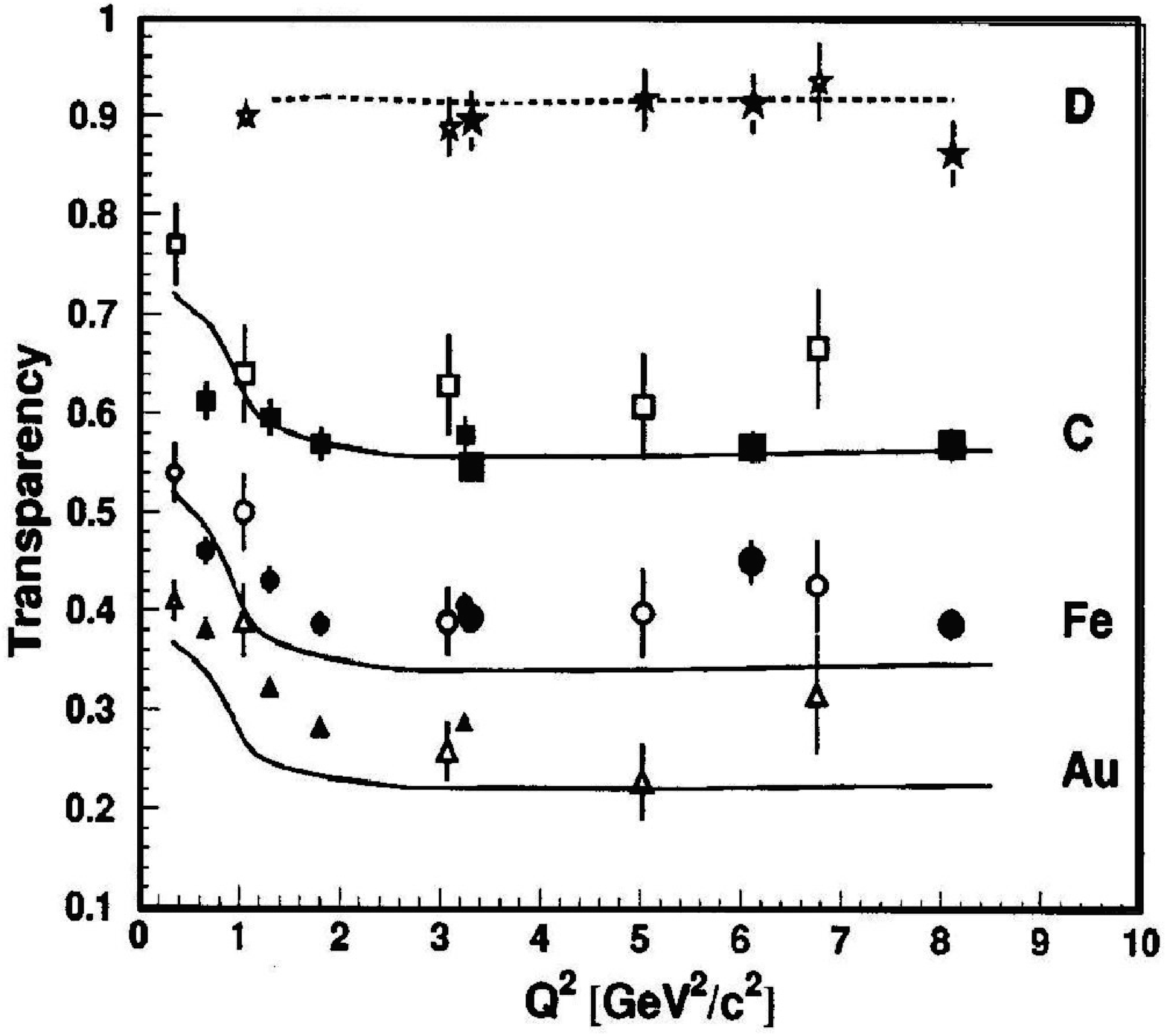}
\caption{(e,e'p) transparencies taken from Reference~\cite{Jlab}.
 The open symbols correspond to the 
SLAC experiments \cite{SLAC1,SLAC2}, and the closed symbols are 
from the more recent results of Jefferson Laboratory \cite{Abbott,Jlab}.}
\label{fig:Garrow_ep}
\end{figure}

As seen in Figure~\ref{fig:Garrow_ep}, these measurements begin with $Q^2$
below 1 $(GeV/c)^2$, and extend to $Q^2$ = 8.1 $(GeV/c)^2$. Above a 
$Q^2$ of $\sim2 (GeV/c)^2$ the measurements are consistent with a constant 
nuclear transparency for the deuterium, carbon, iron, gold targets. Note 
that $Q^2$ = 8.1 $(GeV/c)^2$ corresponds to an incident proton momentum
of 9.5 GeV/c.

The level of nuclear transparency  in the nucleus 
of carbon is predicted by Glauber models to be much larger
for (e,e'p) quasielastic scattering, $T_{ep}$. This is
because of the non-absorptive nature of the two legs of the electron diagram
as they pass through the nucleus. For example, if
$T_{ep}$ for carbon is $\sim0.6$, then one would expect the
corresponding $T_{pp}$ to be $\sim0.2$.  However, note that 
the C(e,e'p) nuclear transparency is not much higher 
than the nuclear transparency for C(p,2p) at its peak.
These experiments reported no significant deviation from the Glauber 
prediction, although the size of the effect would have been
smaller than in the A(p,2p) case.
This may suggest that the phenomena that we observed may be
specific to the QCD dynamics of hard pp collisions.
Both the Ralston-Pire and the Brodsky-deTeramond 
models for energy dependence 
would apply to pp but not to ep scattering 
\cite{Ralston:1988rb,Brodsky:1988xw}.
 Both models do, of course,
contain elements of color transparency that
would be common to ep and pp experiments.

\section{Suggestions for  Future Experiments}
Clearly there remain a number of interesting
investigations involving nuclear 
transparency of protons and other hadrons. 
A revival of the AGS fixed target program \cite{AGS_2000}, or the
construction of the 50-GeV accelerator as part of the J-PARC complex in
Japan \cite{J-PARC}, would provide excellent opportunities to expand 
the range of these nuclear transparency studies.
Some of the remaining questions are:
\begin{itemize}
\item[1.]  What happens at higher incident momentum? 
Does nuclear transparency rise again 
above 20 GeV/c, as predicted in the 
Ralston-Pire picture? \cite{Jain:1996dd}
\item[2.]   A dependent studies in the 12 to 15 GeV/c range; 
will the effective 
absorption cross section continue to fall after
the nuclear transparency stops rising at $\sim9.5$ GeV/c? \cite{Jain:1996dd} 
\item[3.] At the higher energy ranges of these experiments the spin
effects are expected to be greatly diminished. However, they
continue to persist, as shown in both single and double spin
measurements \cite{Saroff:1990gn,Crosbie:1981tb}.  So it is important to
see, in quasielastic scattering inside a nucleus, whether a relatively 
pure pQCD state is selected, and if the spin dependent effects 
are attenuated
\item[4.]  Measurements of nuclear transparency with anti-protons, pions and 
kaons will be informative. These particles have widely 
different cross sections at 
$90^{\circ}_{c.m.}$. For instance, the $pp$ differential cross section 
at $90_{c.m.}^\circ$ is 50-times larger than the $\overline{p}p$ differential
cross section \cite{CW}.  How should this small size
of the $\overline{p}p$ cross section affect the absorption of
$\overline{p}$'s by annihilation?  
\item[5.] The production of exclusively produced resonances
provides a large testing ground for nuclear transparency effects. This is  
especially true for those resonances that allow the determination 
of final state spin orientation, such as 
$\rho$'s or $\Lambda$'s \cite{GB,CW}.  Will the interference terms 
that generate asymmetries disappear for reactions which take
place in the nucleus?
\item[6.] Measurements in light nuclei that determine the probability
of a second hard scatter after the first hard interaction
are an alternative way to study nuclear transparency effects.
With the proper kinematics selected, the probability of the second scatter
is dependent on the state of the hadrons at the first hard interaction
\cite{Frankfurt:1996kk}.
\end{itemize} 

\section{Summary}

Presented here is a summary of the results and interpretations of our $T_{pp}$ 
investigation results, as well as other quasielastic experiments.

\begin{itemize}
\item[1.] Two separate experimental programs, E834 and E850, 
that employed entirely different detectors
and different analysis methods, showed a
consistent and striking energy dependence in the
$T_{pp}$ nuclear transparency for effective momenta from 5 to
 15.8 GeV/c.  This corresponds to
a range in $Q^2$ from 3.9 to 14.0 $(GeV/c)^2$.  
There is a  peak in the nuclear transparency  at
$\sim9.5$ GeV/c.  At the peak, the nuclear transparency is a 
factor of 2 above the maximum expected by the Glauber prediction.  
For 12 GeV/c and above, the nuclear transparency is at or below the
Glauber level.  The probability that these transparencies are
 consistent with a constant is less than 1\%.
The interpretations for the rise and fall of the A(p,2p) transparencies all 
involve the presence of two amplitudes in the pp scattering process
\cite{Ralston:1988rb,Brodsky:1988xw}.

\item[2.]  The A dependent transparencies from E834 provide an independent 
method for measuring the effective cross section as the incident 
momentum is raised. At 5.9 
GeV/c the cross section is 18 $\pm$ 2 mb, while at 10 GeV/c it is 12 $\pm$ 2.6 
mb.

 \item[3.] A number of other measurements have been made with the 
E834 and E850 detectors.
\begin{itemize}
\item[a.]  Measurements on deuterium yielded a nuclear transparency consistent 
with 1.0,
as expected for this small nucleus.
\item[b.] The angular dependence for $T_{pp}$ at 5.9 GeV/c is 
unexpectedly sharp. 
\item[c.] An initial measurement for A$(\pi^+,\pi^+ p)$ has been made and 
indicates that the nuclear transparency for $\pi^+p$ 
is $\sim1.5 \times$ larger than that for pp.
\end{itemize}

\item[4.]  (e,e'p) experiments have shown no energy dependence between  
 $Q^2>\sim2(GeV/c)^2$ and $8.1 (GeV/c)^2$. 
Most theoretical interpretations indicate that, 
unlike the A(e,e'p) case, there are
interferences between the long and short ranged (pQCD scaling) amplitudes
for A(p,2p) scattering.

\item[5.]  The same detectors that measured nuclear transparencies can explore
many other phenomena. This is especially true for measurements
that exploit the abilities of high momentum transfer
quasielastic proton scattering to examine 
the behavior of individual protons in nuclei. These measurements
include nucleon-nucleon correlations. Another investigation is the further
 study of high momentum tails utilizing the enhancement from 
the $s^{-10}$ behavior of pp scattering. 


\end{itemize}

These two A(p,2p) experiments at the AGS have revealed fascinating phenomena 
about nuclear transparency and its possible interpretations with QCD.  Also 
E834 and E850 have shown the usefulness of protons as short-ranged probes of 
nuclear distributions.  We hope that future experiments will expand the 
precision and range of our experiments with protons and other hadrons.

\section{Acknowledgements}

S. Baker, F. J. Barbosa, S. Kaye, M. Kmit, D. Martel, D. Maxam,
J. E. Passaneau, and M.Zilca contributed significantly to the
design, construction, and testing of the detector components. 
In particular, F. Barbosa was responsible for the design of
straw tube amplifiers and readout system.  We are
pleased to acknowledge the assistance of the AGS staff in 
assembling the detector and supporting the experiment.
The experiment benefited  throughout by the diligent
work of our liaison engineers, D. Dayton, J. Mills, and
C. Pearson.  The continuing support of the department
chair, D. Lowenstein, and division head, P. Pile is 
gratefully acknowledged.  This research was supported by 
the U.S.-Israel Binational Science Foundation, the 
Israel Science Foundation founded by the Israel Academy
of Sciences and Humanities, the U.S. National Science Foundation
(Grant No. PHY9501114), and the U.S. Department of Energy 
(Grant No DEFG0290ER40553)

\section{Appendix - Numerical Values}
\nopagebreak
In the following appendix the numerical values of the transparencies
from E850 and E834 are tabulated in Tables I to IV. 
The estimated experimental
systematic error for each specific experiment is given in 
each table of transparencies. Also, there is an overall uncertainty 
of $\pm15$ $\%$ due to theoretical considerations, 
as discussed in Subsection~\ref{subsec:Th_Un}, which 
applies to all the tables.

\begin{table*}[tbp]
\caption{ Table of carbon and deuterium transparencies from E850. 
Column 1 lists
$P_1$,  the incident beam momentum in GeV/c, and 
Column 2 gives $\theta_{c.m.}$ gives
the range of c.m. angles accepted by the detector.  Column 3 lists
$\alpha_1$ and $\alpha_2$, the limits  to $\alpha_0$. 
 $\alpha_0$ is the approximation to the
longitudinal light cone fraction as described
in Subsection IIB. $P_{eff}$ is the effective incident momentum
resulting from the range of $\alpha_0$ given by
$P_{eff}=0.5(\alpha_1+\alpha_2)P_1$. The nuclear transparency ratio of
C to H in the interval $\alpha_1<\alpha_0<\alpha_2$ is
$T_{CH}$.
 The value of the integral, ${\int_{\alpha_1}^{\alpha_2}},$~~ 
$\left(\int_{\alpha_1}^{\alpha_2} d\alpha
N(\alpha) \frac{\frac{d\sigma}{~dt_{pp}}(s(\alpha))}
{\frac{d\sigma}{~dt_{pp}}(s_0)}\right)$, measures the fraction of the
 momentum distribution contained with the limits of $\alpha$ and
 corrects for $s$ dependence of
 $\frac{d \sigma}{{dt}_{pp}}$.  The inverse of ${\int_{\alpha_1}^{\alpha_2}}$
 multiplies $T_{CH}$ to give the nuclear transparency for A(p,2p) quasielastic
 scattering, $T_{pp}$.
 The experiment specific systematic error is ${\pm}4\%$ as discussed in
 Section~\ref{subsec:TRatio}.  In addition, there is an additional overall
 uncertainty due to theoretical considerations.
}
\label{tab:E850_T}
\vspace{0.5cm}
\begin{tabular}{|l|l|l|l|l|l|l|}\hline
$P_1 ~[GeV/c]$
        & $\theta_{c.m.}~[Deg]$
        &$\alpha_0$
        &$P_{eff}~[GeV/c]$
        &$T_{CH}$
        &${\int_{\alpha_1}^{\alpha_2}}$
        &$T_{pp}$
        \\ \hline
\multicolumn{6}{c}{E850 Carbon Data: Leksanov, et al (2001), \cite{E850}}
\\\hline
        \textsf{5.9}
        & \textsf{86.2-90}
        & \textsf{.95-1.05}
        & \textsf{5.9}
        & \textsf{0.071}
        $\pm$ \textsf{0.012}
        & \textsf{0.350}
        & \textsf{0.20} $\pm$ \textsf{0.03}
        \\ \hline
          \textsf{8.0}
        & \textsf{87.0-90}
        & \textsf{.95-1.05}
        & \textsf{8.0}
        & \textsf{0.120}
        $\pm$ \textsf{0.018}
        & \textsf{0.350}
        & \textsf{0.34} $\pm$ \textsf{0.05}
\\\hline
          \textsf{9.1}
        & \textsf{86.8-90}
        & \textsf{.95-1.05}
        & \textsf{9.1}
        & \textsf{0.164}
        $\pm$ \textsf{0.038}
        & \textsf{0.350}
        & \textsf{0.47} $\pm$ \textsf{0.11}
\\\hline
          \textsf{11.6}
        & \textsf{85.8-90}
        & \textsf{.95-1.05}
        & \textsf{11.6}
        & \textsf{0.079}
        $\pm$ \textsf{0.021}
        & \textsf{0.340}
        & \textsf{0.23} $\pm$ \textsf{0.06}
\\\hline
          \textsf{14.4}
        & \textsf{86.3-90}
        & \textsf{.95-1.05}
        & \textsf{14.4}
        & \textsf{0.033}
        $\pm$ \textsf{0.024}
        & \textsf{0.340}
        & \textsf{0.10} $\pm$ \textsf{0.07}
\\\hline
\multicolumn{6}{c}{E850 Carbon Data: for $\alpha > 1$, \cite{Alex_th}}
\\\hline
          \textsf{9.1}
        & \textsf{86.8-90}
        & \textsf{1.05-1.15}
        & \textsf{10.0}
        & \textsf{0.059}
        $\pm$ \textsf{0.015}
        & \textsf{0.11}
        & \textsf{0.53} $\pm$ \textsf{0.15}
\\\hline
          \textsf{11.6}
        & \textsf{85.8-90}
        & \textsf{1.05-1.15}
        & \textsf{12.8}
        & \textsf{0.016}
        $\pm$ \textsf{0.007}
        & \textsf{0.12}
        & \textsf{0.14} $\pm$ \textsf{0.07}
\\\hline
          \textsf{14.4}
        & \textsf{86.3-90}
        & \textsf{1.05-1.15}
        & \textsf{15.8}
        &\textsf{0.007}
        $\pm$ \textsf{0.007}
        & \textsf{0.11}
        & \textsf{0.06} $\pm$  \textsf{0.07}
\\\hline
\multicolumn{6}{c}{E850 Carbon Results: I. Mardor, et al (1998), \cite{IM}}
\\\hline
        \textsf{5.9}
        & \textsf{85.8-90}
        & \textsf{.95-1.05}
        & \textsf{5.9}
        & \textsf{0.054}
        $\pm$ \textsf{0.006}
        & \textsf{0.350}
        & \textsf{0.16} $\pm$ \textsf{0.02}
\\\hline
        \textsf{7.5}
        & \textsf{85.8-90}
        & \textsf{.95-1.05}
        & \textsf{7.5}
        & \textsf{0.072}
        $\pm$ \textsf{0.006}
        & \textsf{0.350}
        & \textsf{0.20} $\pm$ \textsf{0.02}
        \\ \hline
\multicolumn{6}{c}{E850 Deuterium Results: I. Mardor, et al (1988) 
\cite{IM,Israel_th}}

\\\hline
        \textsf{5.9}
        & \textsf{85.5-90}
        & \textsf{.85-1.05}
        & \textsf{5.6}
        & \textsf{-} \textsf{-}
        & \textsf{$\sim$1.0} 
        & \textsf{1.06} $\pm$ \textsf{0.07}
\\\hline
        \textsf{7.5}
        & \textsf{85.5-90}
        & \textsf{.85-1.05}
        & \textsf{7.1}
        & \textsf{-} \textsf{-}
        & \textsf{$\sim$1.0}
        & \textsf{1.10} $\pm$ \textsf{0.10}
        \\ \hline
\end{tabular}
\end{table*}

\begin{table*}[tbp]
\caption{\label{tab:E834_nuc}
  Table of nuclear transparencies for Li, C, Al, Cu, and Pb
from E834 \cite{E834}.  Detector acceptance is
$80^0 < \theta_{c.m.} < 90^0$. The average A of this natural
isotopic abundance targets is listed.
The first three columns have the
same meaning as in Table~\ref{tab:E850_T}. The column labeled $T_{E834}$
gives the transparencies as reported in Reference~\cite{E834}.
The ratio of integrals (see Equation~\ref{eqn:ratio_E850_E834})
 listed in Column 5 corrects
 the approximate nuclear momentum distributions used in E834 with the
 improved nuclear momentum distributions derived
from Reference~\cite{CdA} as employed in
 the analysis of E850.  $T_{pp}$ is the product of $T_{E834}$ with
 the ratio of these integrals.
  The experiment specific systematic error is $\pm5\%$ as discussed in
 Section~\ref{subsec:Kin_E834}.
}


\vspace{0.5cm}
\begin{tabular}{|l|l|l|l|l|l|l|}\hline
$P_1~[GeV/c]$
        &$\alpha_0$
        &$P_{eff}~[GeV/c]$
        &$T_{E834}$
        &$\frac{\int_{pp}}{\int_{E834}}$
        &$T_{pp}$
\\ \hline
      \multicolumn{6}{c}{Lithium(6.9)}
\\ \hline
        \textsf{5.9}
        & \textsf{0.9-1.2}
        & \textsf{6.2}
        & \textsf{0.46}
        $\pm$ \textsf{0.02}
        & \textsf{0.713}
        & \textsf{0.33} $\pm$ \textsf{0.02}
\\ \hline
        \textsf{10.0}
        & \textsf{0.9-1.2}
        & \textsf{10.5}
        & \textsf{0.54}
        $\pm$ \textsf{0.09}
        & \textsf{0.713}
        & \textsf{0.38} $\pm$ \textsf{0.06}
\\ \hline
     \multicolumn{6}{c}{Carbon(12.0)}
\\ \hline
        \textsf{5.9}
        & \textsf{0.9-1.2}
        & \textsf{6.2}
        & \textsf{0.33}
        $\pm$ \textsf{0.02}
        & \textsf{0.736}
        & \textsf{0.24} $\pm$ \textsf{0.02}
\\ \hline
        \textsf{10.}
        & \textsf{0.9-1.2}
        & \textsf{10.5}
        & \textsf{0.43}
        $\pm$ \textsf{0.04}
        & \textsf{0.736}
        & \textsf{0.32} $\pm$ \textsf{0.03}
\\ \hline
        \textsf{12}
        & \textsf{0.9-1.2}
        & \textsf{12.6}
        & \textsf{0.33}
        $\pm$ \textsf{0.10}
        & \textsf{0.736}
        & \textsf{0.24} $\pm$ \textsf{0.07}
\\ \hline
      \multicolumn{6}{c}{Aluminum(27.0)}
\\ \hline
        \textsf{5.9}
        & \textsf{0.9-1.2}
        & \textsf{6.2}
        & \textsf{0.23}
        $\pm$ \textsf{0.15}
        & \textsf{0.715}
        & \textsf{0.16} $\pm$ \textsf{0.01}
\\ \hline
        \textsf{10}
        & \textsf{0.9-1.2}
        & \textsf{10.5}
        & \textsf{0.35}
        $\pm$ \textsf{0.02}
        & \textsf{0.715}
        & \textsf{0.25} $\pm$ \textsf{0.02}
\\ \hline
        \textsf{12}
        & \textsf{0.9-1.2}
        & \textsf{12.6}
        & \textsf{0.20}
        $\pm$ \textsf{0.03}
        & \textsf{0.715}
        & \textsf{0.14} $\pm$ \textsf{0.02}
\\ \hline
      \multicolumn{6}{c}{Copper(63.5)}
\\ \hline
        \textsf{5.9}
        & \textsf{0.9-1.2}
        & \textsf{6.2}
        & \textsf{0.13}
        $\pm$ \textsf{0.02}
        & \textsf{0.713}
        & \textsf{0.09} $\pm$ \textsf{0.02}
\\ \hline
        \textsf{10}
        & \textsf{0.9-1.2}
        & \textsf{10.5}
        & \textsf{0.27}
        $\pm$ \textsf{0.04}
        & \textsf{0.713}
        & \textsf{0.19} $\pm$ \textsf{0.03}
\\ \hline
    \multicolumn{6}{c} {Lead(207.2)}
\\ \hline
        \textsf{5.9}
        & \textsf{0.9-1.2}
        & \textsf{6.2}
        & \textsf{0.058}
        $\pm$ \textsf{0.010}
        & \textsf{0.762}
        & \textsf{0.044} $\pm$ \textsf{0.008}
\\ \hline
        \textsf{10}
        & \textsf{0.9-1.2}
        & \textsf{10.5}
        & \textsf{0.084}
        $\pm$ \textsf{0.025}
        & \textsf{0.762}
        & \textsf{0.064} $\pm$ \textsf{0.019}
\\\hline
\end{tabular}
\end{table*}

\begin{table*}[tbp]
\caption{\label{tab:E834_Al}
  Table of Aluminum  transparencies from E834 for $\alpha \neq 1.0$.
  Detector acceptance is $80^0 < \theta_{c.m.} < 90^0$. Columns
1 to 6 have the same meaning as Tables~\ref{tab:E850_T} and
\ref{tab:E834_nuc} including
the correction of the nuclear momentum distributions from those of
E834 to E850.  Column~7 is the product of $T_{pp}(Al)$ by the
approximate A-dependence
to compare with the corresponding transparencies for Carbon.
The experiment specific systematic error is $\pm5\%$ as discussed in
Section~\ref{subsec:Kin_E834}.
}
\vspace{0.5cm}
\begin{tabular}{|l|l|l|l|l|l|l|l|}\hline
$P_1 ~[GeV/c]$
        &$\alpha_0$
        &$P_{eff} ~[GeV/c]$
        &$T_{E834}$
        &$\frac{\int_{pp}}{\int_{E834}}$
        &$T_{pp}$
        &$T_{pp}\times(\frac{27}{12})^{1/3}$
\\ \hline
        \textsf{5.9}
        & \textsf{.8-.9}
        & \textsf{5.0}
        & \textsf{.18}
        $\pm$ \textsf{0.03}
        & \textsf{.85}
        & \textsf{0.15} $\pm$ \textsf{0.03}
        & \textsf{0.20} $\pm$ \textsf{0.04}
\\ \hline
        \textsf{5.9}
        & \textsf{.9-1.0}
        & \textsf{5.6}
        & \textsf{.25}
        $\pm$ \textsf{0.03}
        & \textsf{.65}
        & \textsf{0.16} $\pm$ \textsf{0.02}
        & \textsf{0.21} $\pm$ \textsf{0.03}
\\ \hline
        \textsf{5.9}
        & \textsf{1.0-1.2}
        & \textsf{6.5}
        & \textsf{.22}
        $\pm$ \textsf{0.04}
        & \textsf{.78}
        & \textsf{0.17} $\pm$ \textsf{0.03}
        & \textsf{0.22} $\pm$ \textsf{0.04}
\\ \hline
        \textsf{10}
        & \textsf{.8-.9}
        & \textsf{8.5}
        & \textsf{.32}
        $\pm$ \textsf{0.04}
        & \textsf{.86}
        & \textsf{0.28} $\pm$ \textsf{0.03}
        & \textsf{0.37} $\pm$ \textsf{0.04}
\\ \hline
        \textsf{10}
        & \textsf{.9-1.0}
        & \textsf{9.5}
        & \textsf{.48}
        $\pm$ \textsf{0.05}
        & \textsf{.65}
        & \textsf{0.31} $\pm$ \textsf{0.03}
        & \textsf{0.41} $\pm$ \textsf{0.04}
\\ \hline
        \textsf{10.}
        & \textsf{1.0-1.2}
        & \textsf{11.0}
        & \textsf{.25}
        $\pm$ \textsf{0.06}
        & \textsf{.81}
        & \textsf{0.20} $\pm$ \textsf{0.05}
        & \textsf{0.26} $\pm$ \textsf{0.07}
\\ \hline
        \textsf{12}
        & \textsf{.8-.9}
        & \textsf{10.2}
        & \textsf{.24}
        $\pm$ \textsf{0.08}
        & \textsf{.85}
        & \textsf{0.20} $\pm$ \textsf{0.07}
        & \textsf{0.26} $\pm$ \textsf{0.09}
\\ \hline
        \textsf{12}
        & \textsf{.9-1.0}
        & \textsf{11.4}
        & \textsf{.2}
        $\pm$ \textsf{0.04}
        & \textsf{.66}
        & \textsf{0.13} $\pm$ \textsf{0.03}
        & \textsf{0.17} $\pm$ \textsf{0.04}
\\ \hline
        \textsf{12.}
        & \textsf{1.0-1.2}
        & \textsf{13.2}
        & \textsf{.12}
        $\pm$ \textsf{0.04}
        & \textsf{.79}
        & \textsf{0.09} $\pm$ \textsf{0.03}
        & \textsf{0.12} $\pm$ \textsf{0.04}
\\\hline
\end{tabular}
\end{table*}


\begin{table*}[tbp]
\caption{\label{tab:pi-p_Tr}
  Table of A$(\pi^+,\pi^+p)$ nuclear transparencies from E834
\cite{JYW}.
 The experiment measured  $T_{\pi^+p}$/$T_{pp}$ ratios, which we have
multiplied by the $T_{pp}$ transparencies of Table~\ref{tab:E834_nuc}
to obtain $(\pi^+,\pi^+p)$ transparencies. The experiment specific
systematic error is $\pm5\%$ as discussed in Section~\ref{subsec:Kin_E834}.
}
\vspace{0.5cm}
\begin{tabular}{|l|l|l|l|l|l|}\hline
$P_1 ~[GeV/c]$
        &$\alpha_0$
        &$T_{pp}$
        &$\frac{T_{\pi^+ p}}{T_{pp}}$
        &$T_{\pi^+ p}$
\\ \hline
       \textsf{Lithium(6.9)}
\\ \hline
        \textsf{5.9}
        & \textsf{0.8-1.2}
        & \textsf{0.33} $\pm$ \textsf{0.014}
        & \textsf{1.10} $\pm$ \textsf{0.15}
         & \textsf{0.36} $\pm$ \textsf{0.05}
\\ \hline
      \textsf{Carbon(12)}
\\ \hline
        \textsf{5.9}
        & \textsf{0.8-1.2}
        & \textsf{0.24} $\pm$ \textsf{0.015}
        & \textsf{1.42} $\pm$ \textsf{0.17}
         & \textsf{0.34} $\pm$ \textsf{0.05}
\\ \hline
      \textsf{Aluminum(27)}
\\ \hline
        \textsf{5.9}
        & \textsf{0.8-1.2}
        & \textsf{0.16} $\pm$ \textsf{0.02}
        & \textsf{1.39} $\pm$ \textsf{0.19}
        & \textsf{0.23} $\pm$ \textsf{0.04}
\\ \hline
        \textsf{10}
        & \textsf{0.8-1.2}
        & \textsf{0.25} $\pm$ \textsf{0.02}
        & \textsf{1.65} $\pm$ \textsf{0.31}
        & \textsf{0.41} $\pm$ \textsf{0.08}
\\ \hline
       \textsf{Copper(63.5)}
\\ \hline
        \textsf{5.9}
        & \textsf{0.8-1.2}
        & \textsf{0.09} $\pm$ \textsf{0.01}
        & \textsf{1.63} $\pm$ \textsf{0.30}
         & \textsf{0.15} $\pm$ \textsf{0.04}
\\ \hline
     \textsf{Lead(207.2)}
\\ \hline
        \textsf{5.9}
        & \textsf{0.8-1.2}
        & \textsf{0.044} $\pm$ \textsf{0.008}
        & \textsf{1.45} $\pm$ \textsf{0.41}
         & \textsf{0.06} $\pm$ \textsf{0.02}
\\\hline
\end{tabular}
\end{table*}

\cleardoublepage

\begin{thebibliography}{10}

\bibitem{IM}
{I. Mardor, et al}.
\newblock {\em Phys.\ Rev. Lett.} {\bf81}, 5085 (1998).

\bibitem{E850}
{A. Leksanov \it{et al.}}
\newblock {\em Phys. Rev. Lett.} {\bf87}, 212301--1 (2001).

\bibitem{E834}
{A. S. Carroll \it{et al.}}
\newblock {\em Phys. Rev. Lett.} {\bf61}, 1698 (1988).

\bibitem{JYW}
J-Y Wu.
\newblock {\em PhD Thesis, Pennsylvania State University}.
\newblock (1992).

\bibitem{Mardor:1998fz}
{Y.~Mardor \it{et al.}}
\newblock {\em Phys. Lett. B} {\bf437}, 257 1998.

\bibitem{Aclander:1999fd}
{J.~Aclander \it{et al.}}
\newblock {\em Phys. Lett. B} {\bf453}, 211  (1999).

\bibitem{Tang}
{A. Tang \it{et al.}}
\newblock {\em Phys. Rev. Lett.} {\bf90}, 042301 (2003).

\bibitem{Malki:2000gh}
{A.~Malki \it{et~al.}}
\newblock {\em Phys. Rev. C} {\bf65}, 015207  (2002).

\bibitem{SpecFn}
{S.~Heppelmann \it{et al.}}
\newblock {\em Phys. Lett. B} {\bf232}, 167 (1989).

\bibitem{SB}
S.~J. Brodsky.
\newblock {\em Proceedings of the XIII International Symposium on
  Multi-particle Dynamics-1982, edited by W. Kittel, W. Metzer, and A. Stergiou
  (World Scientific, Singapore, 1982)}.
\newblock (1983).

\bibitem{MB}
A.~Mueller.
\newblock {\em Proceedings of the XVII Recontre de Moriond, edited by J.Tran
  Thanh Van (Editions Frontieres, Gif-sur-Yvette, France, 1982)}.
\newblock (1982).

\bibitem{RG}
R.~J. Glauber.
\newblock {\em Lectures in Theoretical Physics edited by W.~E.~Britin \it{et
  al}. (Interscience, New York)}.
\newblock (1959).

\bibitem{FSZ}
L.~L. Frankfurt, M.~I. Strikman, and M.~B. Zhalov.
\newblock {\em Phys. Rev. C} {\bf50}, 2189  (1994).

\bibitem{Fermi_mu_rho}
{M.~R.~Adams \it{et al.}}
\newblock {\em Phys. Rev. Lett.} {\bf74}, 1525 (1995).

\bibitem{CERN_mu_rho}
{M.~Arneodo \it{et al}}.
\newblock {\em Nucl. Phys. B}  {\bf429}, 503 (1994).

\bibitem{HERMES_e_rho}
{K.~Ackerstaff \it{et al}}.
\newblock {\em Nucl. Phys. B} {\bf429},  503 (1994).

\bibitem{di-jet}
{E.~M.~Aitala \it{et al}}.
\newblock {\em Phys. Rev. Lett.} {\bf86}, 4773 (2001).

\bibitem{Kycia}
T.~F. Kycia.
\newblock {\em private communication}.

\bibitem{CW}
{C. White \it{et al}}.
\newblock {\em Phys. Rev. D}  {\bf49},  58 (1994).

\bibitem{CLEO_1}
{D. Andrews \it{et al.}}
\newblock {\em Nucl. Instr. and Meth.} {\bf211}, 474  1983).

\bibitem{meng}
W.~Meng.
\newblock {\em Magnetic fields calculated by Opera-3D/Tosca program, Vector
  Fields Lmtd, 24 Bankside, Kidlington, Oxford, OX5 1JE, UK}.

\bibitem{Kmit}
M.~Kmit, M.~Montag, A.~S. Carroll, F.~J. Barbosa, and S.~J. Baker.
\newblock {\em Large Cylindrical Straw Tube Arrays for the EVA Detector, EP\&S
  Informal Report: 91-4}.
\newblock (1991).

\bibitem{Wu}
{J-Y.~Wu \it{et al.}}
\newblock {\em Nucl. Instrum. Meth. A} {\bf349}, 183  (1994).

\bibitem{Fernando}
F.~J. Barbosa.
\newblock {\em EVA 850 Straw Tube Electronics, E850 Internal Report.} (1991).

\bibitem{Ray_Th}
K.~K. Raychaudhuri.
\newblock {\em PhD Thesis, University of Pennsylvania}.
\newblock (1977).

\bibitem{Simon_th}
Simon Durrant.
\newblock {\em PhD Thesis, Pennsylvania State University}.
\newblock (1994).

\bibitem{Yael_th}
Yael Mardor.
\newblock {\em PhD Thesis, Tel Aviv University}.
\newblock (1997).

\bibitem{Israel_th}
Israel Mardor.
\newblock {\em PhD Thesis, Tel Aviv University}.
\newblock (1997).

\bibitem{Alex_th}
Aleksey Leksanov.
\newblock {\em PhD Thesis, Pennsylvania State University}.
\newblock (2000).

\bibitem{Daniel_th}
Daniel Zhalov.
\newblock {\em PhD Thesis, Pennsylvania State University}.
\newblock (2001).

\bibitem{Mancat}
F.~Karl.
\newblock {\em Survey accomplished with ManCat, 3-D theodolite system, Leica
  Geosystems, Heerbrugg, Switzerland}.

\bibitem{CdA}
C~Ciofi degli Atti and S.~Simula.
\newblock {\em Phys. Rev. C} {\bf53}, 1689 (1996).

\bibitem{Frankfurt:2000ty}
L.~Frankfurt, M.~Strikman, and M.~Zhalov.
\newblock {\em Phys. Lett. B} {\bf503}, 73 (2001).

\bibitem{Crosbie:1981tb}
{E.~A. Crosbie \it{et~al.}}
\newblock {\em Phys. Rev. D} {\bf23}, 600 (1981).


\bibitem{BdeT}
S.~J. Brodsky and G.~de~Teramond.
\newblock {\em Phys. Rev. Lett.} {\bf60}, 1924 (1988).

\bibitem{GB}
{G. Blazey \it{et al.}}
\newblock {\em Phys. Rev. Lett.} {\bf55}, 1820 (1985).

\bibitem{BB}
{B. Baller \it{et al.}}
\newblock {\em Phys. Rev. Lett.} {\bf60}, 1118 (1988).

\bibitem{Yaron}
I.~Yaron, L.~Frankfurt, E.~Piasetzky, M.~Sargsian, and M.~Strikman.
\newblock {\em Phys. Rev C} {\bf66}, 024601 (2002).

\bibitem{FLFS}
G.~R. Farrar, H.~Liu, L.~L. Frankfurt, and M.~I Strikman.
\newblock {\em Phys. Rev. Lett.} {\bf61}, 686 (1988).

\bibitem{JM}
B.~K. Jennings and G.~A Miller.
\newblock {\em Phys. Rev. Lett} {\bf69}, 3619 (1992).

\bibitem{Ralston:1988rb}
John~P. Ralston and Bernard Pire.
\newblock {\em Phys. Rev. Lett.} {\bf61}, 1823 (1988).

\bibitem{Hendry:1974ex}
Archibald~W. Hendry.
\newblock {\em Phys. Rev. D} {\bf10}, 2300 (1974).

\bibitem{Brodsky:1988xw}
Stanley~J. Brodsky and G.~F. de~Teramond.
\newblock {\em Phys. Rev. Lett.} {\bf60}, 1924 (1988).

\bibitem{AGS_2000}
S.~Heppelmann.
\newblock {\em Proceedings of the AGS-2000 Workshop, edited by L. Littenberg
  and J. Sandweiss, BNL 52512 Formal Report}.
\newblock (1996).

\bibitem{Moniz}
{E.~J.~Moniz \it{et~al.}}
\newblock 
\newblock {\em Phys. Rev. Lett.} {\bf26}, 445 (1971).

\bibitem{p_exp}
A.~S. Carroll.
\newblock {\em arXiv:hep-ph/0209288.} {2002}.

\bibitem{SH}
S.~Heppelmann.
\newblock {\em Nuc. Phys. B(Proc. Suppl.)} {\bf12}, 159 (1990).

\bibitem{pJ}
P.K. Jain and J.~P. Ralston.
\newblock {\em Phys.\ Rev. D} {\bf48}, 1104 (1993).

\bibitem{ITAN}
{I. Tanihata, \it{et al.}}
\newblock {\em Phys. Lett. B} {\bf100}, 121 (1981).

\bibitem{sig_tot}
Particle~Data Group.
\newblock 
\newblock {\em Phys. Rev. D} {\bf66}, 010001--261 (2002).

\bibitem{SLAC1}
{N.C.R. Makins \it{et al.}}
\newblock {\em Phys. Rev. Lett.} {\bf72}, 1986 (1994).

\bibitem{SLAC2}
{T.G. O'Neill \it{et al.}}
\newblock {\em Phys. Lett. B} {\bf351}, 87 (1995).

\bibitem{Abbott}
{D.Abbott \it{et al.}}
\newblock {\em Phys. Rev. Lett.} {\bf80}, 5072 (1998).

\bibitem{Jlab}
{K. Garrow \it{et al.}}
\newblock {\em Phys. Rev. C} {\bf66}, 044613 (2002).

\bibitem{J-PARC}
{Y. Yamazaki \it{et al.}}
\newblock {\em Joint Project of KEK/JHF and JAERI/NSP, in Proc. of 1999 Part.
  Accelerator Conf.}
\newblock (2000).

\bibitem{Jain:1996dd}
Pankaj Jain, Bernard Pire, and John~P. Ralston.
\newblock {\em Phys. Rept.} {\bf271}, 67 (1996)

\bibitem{Saroff:1990gn}
S.~Saroff et~al.
\newblock {\em Phys. Rev. Lett.} {\bf64}, 995 (1990).

\bibitem{Frankfurt:1996kk}
L.~L. Frankfurt, W.~R. Greenberg, G.~A. Miller, M.~M. Sargsian, and M.~I.
  Strikman.
\newblock {\em Phys. Lett. B} {\bf369},  201 (1996).

\end{thebibliography}

\end{document}